\documentclass[conference]{IEEEtran}
\usepackage{amsmath,amssymb,amsfonts,chemarrow,balance}
\usepackage{graphicx}
\usepackage{pifont}
\newcommand{\cmark}{\ding{51}}%
\newcommand{\xmark}{\ding{55}}%
\let\labelindent\relax
\usepackage[inline]{enumitem}
\usepackage{mathenv}
\usepackage[english]{babel}
\usepackage{breqn}
\newcommand{\BfPara}[1]{{\noindent\bf#1.}\xspace\xspace}
\usepackage{url}
\usepackage{subfigure}
\usepackage{multirow}
\newcommand{\vs}[1]{{\vspace{-#1mm}}}

\usepackage[linesnumbered,ruled]{algorithm2e}
\usepackage{algpseudocode}
\usepackage{booktabs}
\usepackage{graphics}
\usepackage{epsfig}
\usepackage{listings}
\usepackage{rotating}
\usepackage{amsthm}
\usepackage{hyperref}
\usepackage{xcolor}
\usepackage{upquote}
\usepackage{listings}
\usepackage{xspace}
\usepackage{setspace}
\usepackage{caption}

\newcommand{\note}[1]{}    
\newcommand{\slx}{{\em Selenium}\xspace}
\newcommand{\ch}{{\em Coinhive}\xspace}
\newcommand{\js}{{\em JavaScript}\xspace}
\newcommand{\pl}{{\em Plato}\xspace}

\newcommand{\bcc}{{blockchain}\xspace}

\newcommand{\etc}{{etc.}\xspace}

\newcommand{\cc}{{cryptocurrency}\xspace}
\newcommand{\Cc}{{Cryptocurrency}\xspace}

\newcommand{\cj}{cryptojacking\xspace}
\newcommand\JSONnumbervaluestyle{\color{blue}}
\newcommand\JSONstringvaluestyle{\color{red}}

\newcommand{\etal}{{\em et al.}\xspace}

\colorlet{punct}{red!60!black}
\definecolor{background}{HTML}{ffffff }
\definecolor{delim}{RGB}{20,105,176}
\colorlet{numb}{magenta!60!black}
\definecolor{light-gray}{gray}{0.95}
\definecolor{darkgray}{rgb}{0.4, 0.4, 0.4}
\definecolor{editorGray}{rgb}{0.95, 0.95, 0.95}
\definecolor{editorOcher}{rgb}{1, 0.5, 0} 
\definecolor{editorGreen}{rgb}{0, 0.5, 0} 
\definecolor{orange}{rgb}{1,0.45,0.13}      
\definecolor{olive}{rgb}{0.17,0.59,0.20}
\definecolor{brown}{rgb}{0.69,0.31,0.31}
\definecolor{purple}{rgb}{0.38,0.18,0.81}
\definecolor{lightblue}{rgb}{0.1,0.57,0.7}
\definecolor{lightred}{rgb}{1,0.4,0.5}

\lstdefinelanguage{JavaScript}{
  morekeywords={typeof, new, true, false, catch, function, return, null, catch, switch, var, if, in, while, do, else, case, break},
  morecomment=[s]{/*}{*/},
  morecomment=[l]//,
  morestring=[b]",
  morestring=[b]'
}
\SetKwFor{For}{for (}{) $\lbrace$}{$\rbrace$}
\lstdefinelanguage{HTML5}{
  language=html,
  sensitive=true,   
  alsoletter={<>=-:},    
  morecomment=[s]{<!-}{-->},
  tag=[s],
  otherkeywords={
  >,
    <!DOCTYPE,
  </html, <html, <head, <title, </title, <style, </style, <link, </head, <meta, />,
    </body, <body,
    </div, <div, </div>, 
    </p, <p, </p>,
    </script, <script,
  <canvas, /canvas>, <svg, <rect, <animateTransform, </rect>, </svg>, <video, <source, <iframe, </iframe>, </video>, <image, </image>, <header, </header, <article, </article
  },
  ndkeywords={
  =,
  charset=, src=, throttle:, id=, width=, height=, style=, type=, rel=, href=,
  fill=, attributeName=, begin=, dur=, from=, to=, poster=, controls=, x=, y=, repeatCount=, xlink:href=,
  margin:, padding:, background-image:, border:, top:, left:, position:, width:, height:, margin-top:, margin-bottom:, font-size:, line-height:,
  transform:, -moz-transform:, -webkit-transform:,
  animation:, -webkit-animation:,
  transition:,  transition-duration:, transition-property:, transition-timing-function:,
  }
}

\lstdefinestyle{htmlcssjs} {%
    backgroundcolor=\color{light-gray},
    basicstyle=\footnotesize\ttfamily,
    showstringspaces=false,
    frame = lines, 
    breaklines=true,
    showstringspaces =false,
    keywords = {false,true},
  backgroundcolor=\color{light-gray},
  identifierstyle=\color{black},
  keywordstyle=\color{blue},
  ndkeywordstyle=\color{black},
  stringstyle=\color{red}\ttfamily,
  commentstyle=\color{brown}\ttfamily,
  language=HTML5,
  alsolanguage=JavaScript,
  alsodigit={.:;},  
  tabsize=2,
  showtabs=false,
  showspaces=false,
  showstringspaces=false,
  extendedchars=true,
  breaklines=true,
  literate=%
  {Ã}{{\"O}}1
  {Ã}{{\"A}}1
  {Ã}{{\"U}}1
  {Ã}{{\ss}}1
  {ÃŒ}{{\"u}}1
  {Ã€}{{\"a}}1
  {Ã¶}{{\"o}}1
}

\DeclareCaptionFormat{listing}{{
    \parbox{0.5\textwidth}{\hspace{1pt}#1#2#3}
  }}
\lstdefinestyle{json}
{ backgroundcolor=\color{light-gray},
    basicstyle=\footnotesize\ttfamily,
    showstringspaces=false,
    breaklines=true,
    frame=lines,
  showstringspaces    =false,
  keywords            = {false,true},
  commentstyle= \itshape\color{codegreen},
  alsoletter          =0123456789.,
  morestring          = [s]{"}{"},
  stringstyle         = \ifcolonfoundonthisline\JSONstringvaluestyle\fi,
  MoreSelectCharTable =%
    \lst@DefSaveDef{`:}\colon@json{\processColon@json},
  keywordstyle        = \ttfamily\bfseries,
}

\SetKwRepeat{Do}{do}{while}%

\def\equationautorefname~#1\null{(#1)\null}
\newif\ifcolonfoundonthisline
\makeatletter

\newcommand\processColon@json{%
  \colon@json%
  \ifnum\lst@mode=\lst@Pmode%
    \global\colonfoundonthislinetrue%
  \fi
}

\lst@AddToHook{Output}{%
  \ifcolonfoundonthisline%
    \ifnum\lst@mode=\lst@Pmode%
      \def\lst@thestyle{\JSONnumbervaluestyle}%
    \fi
  \fi
  \lsthk@DetectKeywords%
}

\lst@AddToHook{EOL}%
  {\global\colonfoundonthislinefalse}
\makeatother

\hypersetup{
	plainpages=false,
	colorlinks,
	urlcolor=blue,
	linkcolor=blue,
	citecolor=blue,
	bookmarksnumbered
}

\begin{document}

\title{End-to-End Analysis of In-Browser Cryptojacking}

\author{

\IEEEauthorblockN{{Muhammad Saad}\\
University of Central Florida\\
 saad.ucf@knights.ucf.edu} 
\and 
\IEEEauthorblockN{{Aminollah Khormali}\\
University of Central Florida\\
 aminkhormali@knights.ucf.edu} 
\and
\IEEEauthorblockN{{Aziz Mohaisen}\\
University of Central Florida\\
 mohaisen@cs.ucf.edu} 
}



\maketitle

\begin{abstract}
In-browser cryptojacking involves hijacking the CPU power of a website's visitor to perform CPU-intensive cryptocurrency mining, and has been on the rise, with 8500\% growth during 2017. While some websites advocate cryptojacking as a replacement for online advertisement, web attackers exploit it to generate revenue by embedding malicious cryptojacking code in highly ranked websites. Motivated by the rise of cryptojacking and the lack of any prior systematic work, we set out to analyze malicious cryptojacking statically and dynamically, and examine the economical basis of cryptojacking as an alternative to advertisement.  For our static analysis, we perform content-, currency-, and code-based analyses. Through the content-based analysis, we unveil that cryptojacking is a wide-spread threat targeting a variety of website types. Through a currency-based analysis we highlight affinities between mining platforms and currencies: the majority of cryptojacking websites use Coinhive to mine Monero. Through code-based analysis, we highlight unique code complexity features of cryptojacking scripts, and use them to detect cryptojacking code among benign and other malicious JavaScript code, with an accuracy of $\approx$96.4\%. Through dynamic analysis, we highlight the impact of cryptojacking on system resources, such as CPU and battery consumption (in battery-powered devices); we use the latter to build an analytical model that examines the feasibility of cryptojacking as an alternative to online advertisement, and show a huge negative profit/loss gap, suggesting that the model is impractical. By surveying existing countermeasures and their limitations, we conclude with long-term countermeasures using insights from our analysis.
\end{abstract}

\section{Introduction}\label{sec:introduction}
Recently, \bcc-based cryptocurrencies have emerged as an innovation in distributed systems, enabling a transparent and distributed storage of transactions. To prevent abuse and improve trustworthiness in cryptocurrencies, various proof mechanisms, such as the Proof-of-Work (PoW) and Proof of Stake (PoS), are used. In Bitcoin, one of the most prominent blockchain-based cryptocurrencies, for example, PoW is used to embed  trustworthiness in the system. In Bitcoin, new coins are mined by individual miners through extensive hash operations, which are then verified by distributed nodes in a peer-to-peer (P2P) network. However, the use of PoW in Bitcoin has led to abuse: an adversary may employ various techniques to abuse public resources for mining purposes and to perform extensive hash calculations at no or low cost. 

One such technique that has emerged recently is called {\em cryptojacking}, which involves outsourcing hash calculations in PoW-based cryptocurrencies. Cryptojacking is the use of system resources of a target device to compute hashes and make profit out of mining without the consent of the target device's owner. Conventional \cj involved installation of a software binary on a target machine that secretly solved PoW and communicated the results to a remote server~\cite{Scott_18}. Such conventional \cj required user permission to download the software and a persistent Internet connection to communicate the PoW result to the adversary or a {\em dropzone} server controlled by him. However, conventional \cj proved infeasible for several reasons. First, not all devices have a persistent Internet connection when needed to send PoW results; PoWs are time sensitive, and if not sent immediately after being solved they become easily outdated. Secondly, antivirus companies can easily identify binaries used for \cj and detect them~\cite{Zuckerman_18}. Finally, this form of attack requires an infection vector, whereby users enable the attack by mistakenly installing the \cj binaries on their machines. 

A recent form of in-browser \cj that does not suffer from those issues has emerged. In-browser \cj does not require installing binaries, or authorization from users to operate the system. In-browser \cj instances use \js code to compute PoW in web browser and transmit the PoW to a remote {\em dropzone} server~\cite{Slm_18,Cimpanu_18}. As such, and since they are shielded in the browser's process, they are undetected by the antivirus scanners. Moreover, mining during web browsing  ensures uninterrupted transmission of PoW over a persistent in-place Internet connection. 

\begin{figure}[t]
\centering
\includegraphics[width=0.4\textwidth]{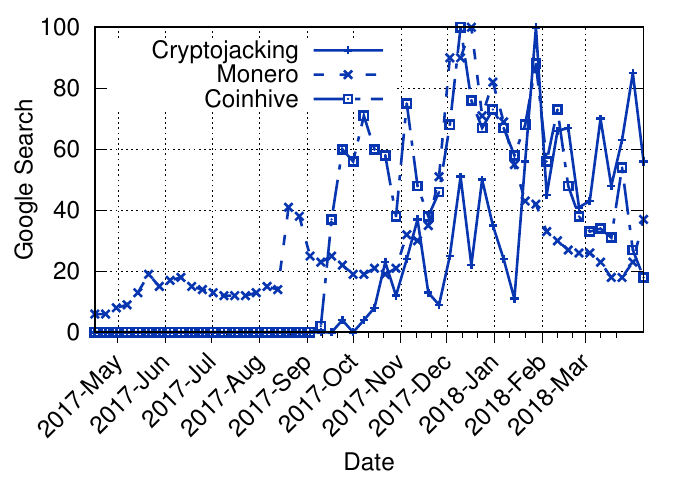}
\caption{Google search trends for Cryptojacking, Monero, and Coinhive over the past 10 months. The results have been normalized in the range [0-100].}
\label{fig:googlesearch}
\vs{8}
\end{figure}

\begin{figure*}[ht]
\centering
		\subfigure[Cryptojacking\label{fig:cryptojacking}] {\includegraphics[width=0.3\textwidth]{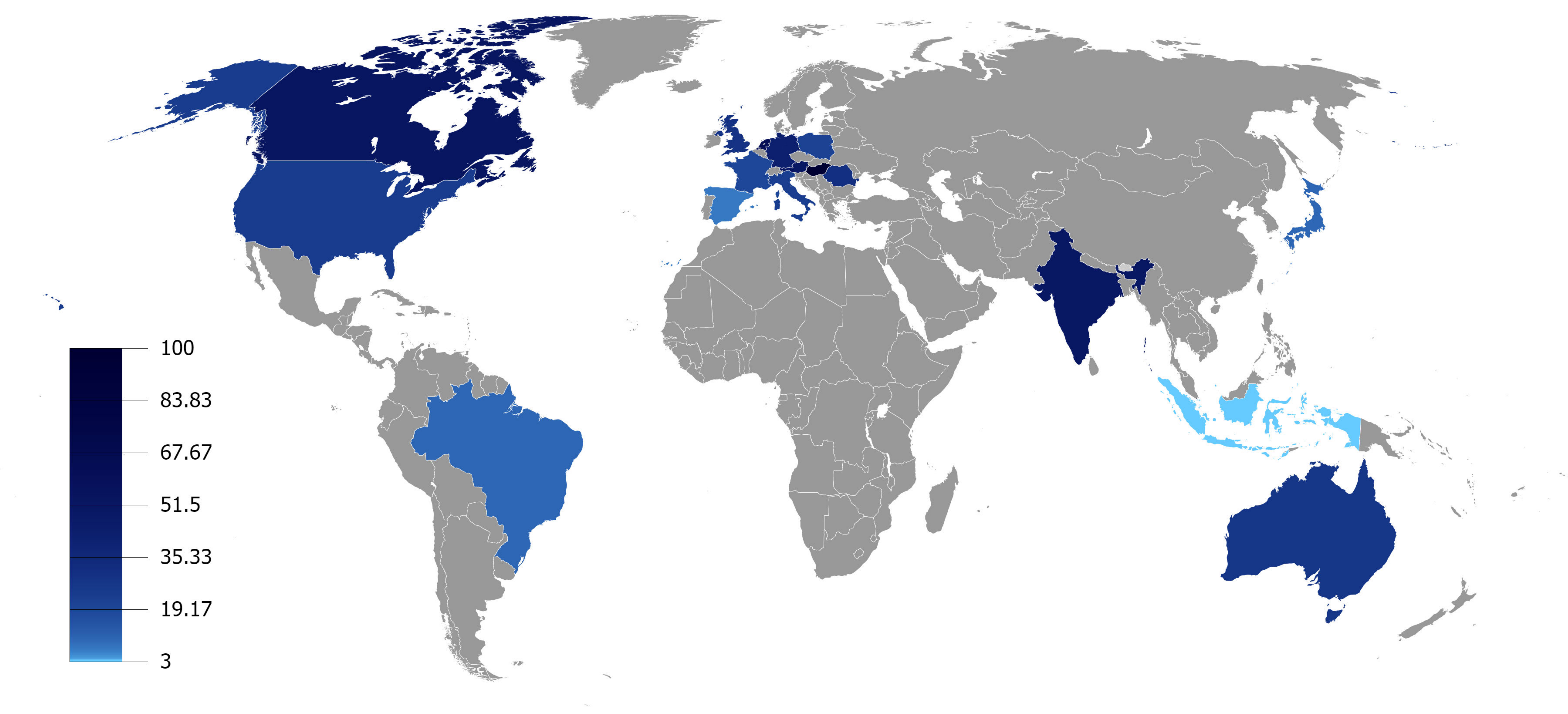}}
		\subfigure[Coinhive \label{fig:coinhive}] {\includegraphics[width=0.3\textwidth]{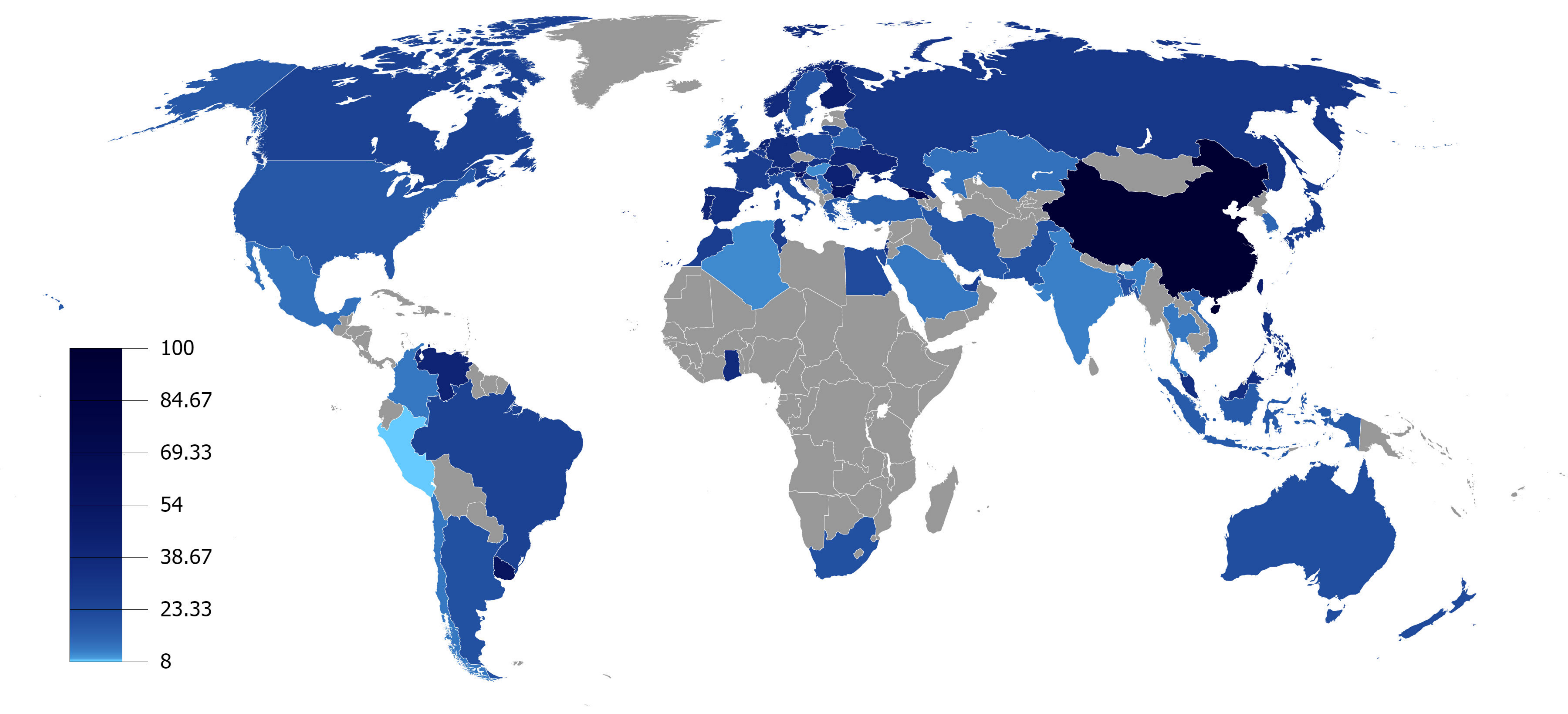}}
		\subfigure[Monero \label{fig:monero}] {\includegraphics[width=0.3\textwidth]{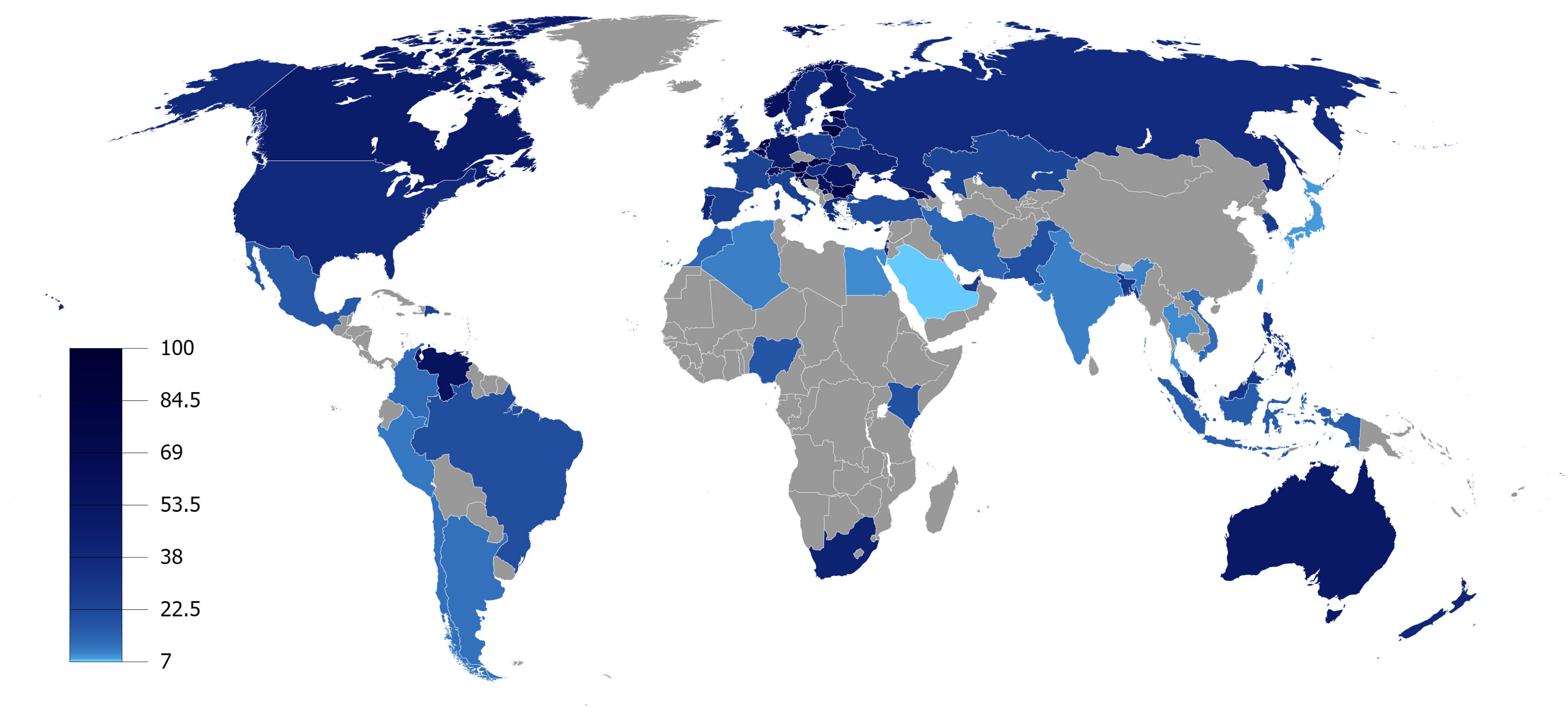}}
\caption{Heatmap of the global distribution of Google searches for each term. Notice that US is the most prevalent country in all three search results. Moreover there is more similarity in the search for Coinhive and Monero.}\label{fig:heatmap}
\vs{4}
\end{figure*}

Initially intended for good use as an alternative revenue source to online advertisement~\cite{kerbs}, in-browser \cj was made easy by online services such as \ch~\cite{coinhive}, which provided \js templates for \cj. \ch provides scripts to mine Monero, a \cc that is hard to trace, and to reward miners based on the aggregated hashes they contribute.  Google terms search reports for ``Cryptojacking'', ``Monero'', and ``Coinhive'' from May 2017 to March 2018 demonstrate the increasing interests in \cj as a global phenomenon, as shown in~\autoref{fig:googlesearch} (and detailed in \autoref{fig:heatmap}). This rise has coincided with the rise in malicious use of in-browser \cj: more than 32,000 websites running \ch scripts, many of which are the result of compromise and \ch scripts injection, are reported~\cite{kerbs}.

In-browser \cj serves as an attack avenue for hackers who inject malicious \js code into popular websites without the knowledge of website owners and mine \cc for themselves. This is known as a \cj attack, and become a major problem recently. According to Symantec's latest Internet Security Threat Report (ISTR), \cj attacks on websites rose by 8500\% during 2017~\cite{Mathur_18,Singh_18}. In February 2018, a major \cj attack hit more than 4000 websites across the world including the websites of US Federal Judiciary and the UK National Health Service (NHS)~\cite{condliffe_18}. Also in February 2018, Tesla became the victim of a \cj attack in which attackers hijacked Tesla cloud and deployed their own \cj code~\cite{Rayome_18}. After such unusual incidents, UK's National Cyber Security Centre (NCSC) indicated \cj as a ``significant threat'' in its latest cyber security report~\cite{de_18,ncsc_18}. 

The use of \cj as a replacement to advertisement also has witnessed a great debate. For example, some popular websites such as ``The Pirate Bay'', among others, started using \cj as a revenue substitute to online advertisement~\cite{Shaikh_17,Ernesto_17,Jones_2017}. The Pirate Bay website later disclosed to its users that it will be using CPU cycles of the visitors in exchange for ads-free web browsing, garnering users approval. As some other websites started using \cj as a revenue generation-mechanism, further debate was sparked surrounding the ethics of using \cj~\cite{Zuckerman_2018}, and the absence of user consent. Furthermore, it was later observed that the continuous CPU-intensive mining, especially on battery-powered devices, has resulted in the quick drainage of those devices, adding a new variable to the debate of whether \cj is a good alternative to online advertising. 


Motivated by these fast-paced and recent events, we carry out the first in-depth study on in-browser \cj and its effects on the website visitors and their devices. We start by analyzing and characterizing more than 5,700 websites that have \cj scripts in them. We then explore both static and dynamic analysis tools to understand the behavioral traits of in-browser \cj scripts towards their detection. Using various features extracted through this analysis, we build a clustering scheme that is used to for detecting \cj scripts among benign scripts, as well as other malicious types of \js codes. We also measure the impact of in-browser \cj on user devices in terms of CPU usage and battery drainage. Finally, in examining the feasibility of \cj as an alternative to online advertisement, we conduct an in-depth end-to-end analysis that considers the implications of such an alternative on both users and websites.

\BfPara{Contributions and Roadmap}
We make the following major contributions:
\begin{enumerate*}
    \item Using more than 5,700 websites with \cj scripts, we conduct the first in-depth analysis and characterization of \cj in the wild, highlighting categories and affinities, including sectors, top-level domains, etc. (\textsection\ref{sec:prelim}). 
    \item Using the same dataset, we conduct static analysis of the \cj scripts, to highlight distributions of \cc used in \cj and code (script) complexity analysis (\textsection\ref{sec:static}). As an application of our static analysis, using code complexity features we built an unsupervised clustering system that automatically identifies \cj, malicious, and benign scripts (\textsection\ref{sec:clustering}).  A reference-based (using ground truth) evaluation of our clustering algorithm yielded an accuracy of $\approx96\%$. 
    \item Using the same dataset, we performed dynamic analysis to highlight the unique characteristics of process usage, battery usage, and dynamically generated data analysis through WebSocket inspection of \cj scripts (\textsection\ref{sec:dynamic}). 
    \item As an application of our dynamic analysis, we explore the economic arguments made for \cj as an alternative to online advertisement, and build an analytical model to estimate the cost of \cj to the users as well as the gain to websites conducting \cj (\textsection\ref{sec:economic}). We supplement this analysis by contrasting it with the existing online advertisement model. We show the economical model of \cj is impractical for benign use, and unprofitable for malicious use. 
    \item We explore the limitations of existing countermeasures and suggest more robust defense techniques to address in-browser \cj using our static and dynamic analysis insights (\textsection\ref{sec:counter}). 
\end{enumerate*}

Additionally, the rest of this paper includes a background in \textsection\ref{sec:background}, the related work in~\textsection\ref{sec:rw}, discussion in~\textsection\ref{sec:discussion}, and concluding remarks in~\textsection\ref{sec:conclusion}, respectively. 


\section{Background}\label{sec:background}
In this section, we review the preliminaries of this work, including an introduction to \cc, the mining process, and \cj. We then outline the problem statement and motivation and data collection. 


\subsection{Blockchain-based Cryptocurrencies}\label{sec:cryptos}
In 2009, the first blockchain-based digital currency ``Bitcoin'' was introduced by Satoshi Nakamoto~\cite{nakamoto2008bitcoin} that involved exchange of transactions without the use of a central authority. In Bitcoin, the role of the trusted central authority was replaced by a transparent and tamper-proof public blockchain that acted as a public ledger to maintain the records of transactions. The consensus in the decentralized peer-to-peer Bitcoin network was augmented by cryptographically secure algorithm known as the proof-of-work (PoW). Bitcoin remained the only \cc for two years after which several more digital currencies joined the market. As of today, there are more than 5000 cryptocurrencies have been introduced in the market~\cite{atozforex} with more than 5.8 million active users~\cite{hileman2017global}. Bitcoin is leading the \cc market with a 58\% market share, or $\approx$\$4.9 Billion USD trade volume and more than 12,000 transactions per hour~\cite{bitcoinnews_2017}. 
Towards the end of 2016, the price of 1 bitcoin was a little under \$1000 USD and during 2017 it witnessed exponential growth rising to a market price of \$19,000 USD~\cite{blockexplorer}. Some other notable cryptocurrencies that make use of public blockchain are Ethereum, Litecoin, Ripple, Monero, and Dash. 

\subsection{Mining in Cryptocurrencies}\label{sec:mining}
The key operations in every \cc involve exchange of transactions among peers, the mining of transactions in blocks, and publishing those blocks. Computing a valid block results in the generation of new coins in the system..

However, computing a valid block is a non-trivial process in which miners have to solve mathematical challenges and provide a PoW for their solutions. In Bitcoin, PoW involves finding a \textit{nonce} that, when hashed with the data in the block, produces a hash value less than the target threshold set by the system. The target is a function of network difficulty and is denoted by a 256-bit unsigned integer that is encoded in a 32-bit ``compact'' form and stored in the block header. In the process of solving the challenge, miners spend effort and in return get rewarded with 12.5 bitcoins for each valid PoW. As more miners join, the hash power of the network and the probability of computing a block increase. To keep the average block computation time within the fixed range, the network's difficulty is adjusted every two weeks (2016 blocks).

In~\autoref{equ:prob}, we show how the block computation time, $T(B)$, is affected by the hashing rate, $H_r$, the {\em target}, $Target$, the probability of finding a block, $P_r(B)$, and the average number of hashes required to solve the target, $H$. To keep  $T(B)$ in a fixed range (10 minutes), as the $H_r$ increases, the target value is adjusted to keep $P_r(B)$ constant. 

\begin{align} 
  \label{equ:prob}  P_r(B)  = \frac{Target}  {2^{256}},  H = \frac{1}{P_r(B)}, T(B) = \frac{1}{P_r(B)\times H_r}
\end{align}

To maximize mining reward, multi-homed mining pools---all participants collaboratively compute hashes based on the hash power of their machines---have emerged. When a block is computed, the rewards are distributed among the participants based on their contribution towards the produced hashes. Mining pools enable even ordinary users with limited mining hardware to effectively participate in the mining process. As a result of this paradigm, there has been an exponential growth in the aggregate hash rate of the cryptocurrencies as more people have shown interest in mining. 

\subsection{Cryptojacking}\label{sec:cj} 
Generally, attackers utilize two main strategies for unauthorized use of a victim's machine to mine digital currencies through \cj:  by installing a binary on the  machine, or by using an in-browser script. The first one loads the mining code on the victim's machine as a stand-alone binary (or an infection of a binary). As such, it requires information about the target machine including its operating system and hardware constructs. For example a malicious \cj binary developed for Windows cannot be executed on Linux. However, the second strategy is platform agnostic, the \cj~\js is executed upon loading the website in victim's browser. In both cases, the mining code works in the background while the unaware victim is using his machine. The focus of this paper is the latter type, which we highlight at length below. In this rest of this paper, we will refer to the abuse case of \cj, whereby an adversary injects \cj scripts to mine cryptocurrencies.

\subsubsection{In-Browser Cryptojacking} \label{sec:brcj}
In-browser \cj is done by injecting a \js code in a website, allowing it to hijack the processing power of a visitor's device to mine a specific \cc. Generally, \js is automatically executed when a website is loaded. Upon visiting a website with \cj code, the visiting host starts a mining activity, by becoming part of a \cj mining pool.  A key feature of in-browser \cj is being platform-independent: it can be run on any host, PC, mobile phone, tablet, etc., as long as the web browser running on this host has \js enabled in it. \js, however, is one of the most popular web languages and, by default, is enabled in most major browsers. Furthermore, in-browser \cj allows for mining at-scale without requiring any custom hardware: as more visitors visit the website with \cj scripts, more processing power is available for mining.

\subsubsection{Cryptojacking as a Replacement to Advertisement} \label{sec:adcj}
An ongoing debate sparked in the community for whether \cj can serve as a replacement to online advertisement. Those advocating the approach have pointed out that users providing their CPU power to a website for mining can use the website without viewing online advertisements. Towards that, some websites, including the aforementioned `The Pirate Bay'', started using \cj as a revenue substitute for online advertisements~\cite{Shaikh_17,Ernesto_17,Jones_2017} and become ``ads-free operation''. However, a counter argument to this model is the claimed to be the excessive abuse of the \cj website to the visitor's CPU resources. In-browser \cj scripts will not only run in the background without a user consent, but would also drain batteries in battery-powered platforms, would indirectly affect the user experience, and by locking the CPU power and not allowing him to use other applications.

\section{Dataset and Preliminary Analysis}\label{sec:prelim}
With the objective of this work, as stated earlier, being the characterization, analysis, and detection of in-browser \cj, as well as testing the economical argument for the \cj as an alternative to online advertisement, we proceed by outlining the data collection procedure we followed and basic characteristics. In the subsequent two sections, we outline the static analysis and dynamic analysis we conducted to uncover \cj scripts. 

\subsection{Data Collection} \label{sec:data}
We assembled a data set of \cj websites published by  Pixalate~\cite{Loechner17} and Netlab 360~\cite{Netlab360}. Pixalate is a network analytics company that provides data solutions for digital advertising and research. In Nov. 2017, they collected
a list of 5,000 cryptojacking websites that were actively stealing visitor’s processing power to mine cryptocurrency. We obtained that list of cryptojacking websites from Pixalate. Netlab 360 (Network Security Research Lab at 360) is a data research platform that provides a wide range of datasets spanning Domain Name Servers (DNS) and Distributed Denial-of-Service (DDoS) attacks. From Netlab 360, we obtained  700 \cj websites, released on Feb 24, 2018. 

The top-level domain (TLD) distribution of the combined dataset, including the TLD type (generic, new, or country-level) and the corresponding  percentage, is shown in~\autoref{tab:pixalate}. While, unsurprisingly, .com and .net occupy the first and second spot of the top-10 TLDs represented in the dataset, with a combined total of 40.3\% of the websites belong to them, country-level domains have a significant presence, with countries such as Slovenia, Russia, and Brazil well represented in the dataset. New-gTLDs were also present in the top-10 gTLDs, with .site having $\approx$2.0\% of the sites.

In the Pixalate's dataset, 6 websites were found in the Alexa top-5000 websites and 13 were among the Alexa top-10000 websites. Among the \cj site, 68.3\% did not have a privacy policy, while 56.8\% websites had no ``terms and conditions'' statement, and  49.3\% did not have both the privacy policy and the terms and conditions. This indicates that the majority of those websites could not formally, through those statements, inform their visitors regarding the usage of their processing resources for mining cryptocurrencies, where \cj is used  instead of online advertisement.


 During our analysis we also observed that 11\% of the websites in our dataset had stopped \cj, due to key revocation by the server, removal of the code from the website, or the closure of websites. We exclude them from our analysis.

\begin{figure*}
\centering
\includegraphics[width=17.6cm]{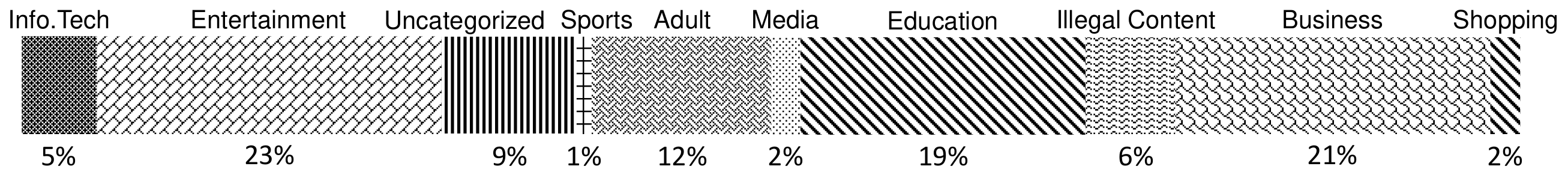}
\caption{Categorization of websites based on the main topic of their content. Notice that most websites belong to Entertainment, Business, and Education. A sizable chunk (12\%) belonged to the Adult category. }
\label{fig:content}
\centering
\end{figure*}

\begin{table}[t]
\centering
\caption{Distribution of cryptojacking websites with respect to top-level domains in our dataset. }
\label{tab:pixalate}

\begin{tabular}{|c|l|l|l|l|}
\hline
{\textbf{Rank}} & {\textbf{TLD}} & {\textbf{Type}} & {\textbf{Sites}} & {\textbf{Sites\%}} \\ \hline
1 & .com                                                   & generic                                                   & 1945                                        & 34.1\%                                      \\ \hline
2 & .net                                                   & generic                                                   &359                                         & 6.2\%                                       \\ \hline

3 & .si                                                    & country                                                   & 358                                         & 6.2\%                                       \\ \hline
4 & .online                                                & generic                                                   & 349                                         & 6.1\%                                       \\ \hline

5 & .ru                                                    & country                                                   &242                                         & 4.2\%                                       \\ \hline
6 & .org                                                   & generic                                                   &191                                        & 3.3\%                                       \\ \hline
7 & .sk                                                    & country                                                   &169                                         & 2.9\%                                       \\ \hline
8 & .info                                                  & generic                                                   &169                                         & 2.9\%                                       \\ \hline

9 & .br                                                    & country                                                   &157                                         & 2.7\%                                       \\ \hline

10 & .site                                                  & new                                                   &116                                         & 2.0\%                                       \\ \hline

11 & others                                                 & {---}                                                                                                & 1648                                        & 28.8\%                                      \\ \hline \hline

\textbf{Total}  &   {---}                                              & {---}  & \textbf{5703}                                        & \textbf{100\%}                                      \\ \hline
\end{tabular}
\end{table}

\subsection{Methodology} \label{sec:method}
After gathering the dataset, we perform static and dynamic analysis of the \cj~\js code. In the static analysis, we categorize the websites based on content and the currency they mine during \cj. We extract the \cj code and develop code-based features to examine their properties. We compare them, using those static properties, with malicious and benign \js code. We use standard code analyzers to extract program specific features. 

In our dynamic analysis, we explore the CPU power consumed by \cj websites and its effects on the user devices. We run test websites to mimic \cj websites and carry out a series of experiments to validate our hypothesis. For our experiments, we use Selenium-based scripts to automate browsers and various end host devices, including Windows and Linux operated laptops and an Android phone, to monitor the effect of \cj under various operating systems and hardware architectures.  For website information, we use services provided by Alexa and SimilarWeb to extract information regarding websites ranking, volume of traffic, and the average time spent by visitors on those websites~\cite{Similarweb-18}. 

\section{Static Analysis} \label{sec:static}
For static analysis, we pursue three directions:  content-\, currency-, and code-based analysis. Content-based categorization provides insights into the nature of websites used for \cj activities, while the currency-based categorization shows the distribution of service providers and platforms providing \cj templates for those websites. Finally, the code-based analysis provides insight into the complexity of the \cj scripts, using various code complexity measures from the literature.  

\subsection{Content-based Categorization} \label{sec:contentcat} For a deeper insight into their usage, it is important to understand what kind of websites have \cj scripts in them. To this end, and as a first step, we categorized the websites based on their contents into various categories using the {\em WebShrinker} website URL categorization API. {\em WebShrinker} assigns categories to websites based on the main usage of those websites using their contents. The results are shown in \autoref{fig:content}. As it can be seen in \autoref{fig:content}, miners have utilized a wide range of categories for in-browser \cc mining, including education, business, entertainment, \etc Notice in~\autoref{fig:content}, some websites are categorized as ``Illegal Content.'' These websites are mostly torrent websites that serve illegal copies of movies and software. Moreover, 19\% websites were categorized as ``Education'' which can be attributed to the exploitation of trust by adversaries behind \cj, since educational sites are highly trusted by their visitors~\cite{ZarrasKSHKV14}. 

\subsection{Currency-based Categorization} To understand the \cj ecosystem, it is critical to find out what cryptocurrencies are typically being mined through in-browser \cj. Therefore, we inspected the websites' scripts to extract information about the platforms and cryptocurrencies.
From our dataset we found that there were eight platforms providing templates to mine two types of cryptocurrencies namely, Monero and JSEcoin. In~\autoref{tab:currencies}, we provide details about the eight platforms and their respective mining \cc. As a result, we found that a very large proportion of the websites ($\approx$81.57\%) use \ch~\cite{coinhive} platform to mine Monero \cc~\cite{Monero}, which is one of the few cryptocurrencies that supports in-browser mining. We found that $\approx$86.37\% of the websites in our dataset are mining Monero \cc through seven platforms. In addition, $\approx$2.61\% of the websites are using the JSEcoin platform~\cite{JSEcoin}, which is responsible for mining the JSEcoin \cc. 

Although PoW-based cryptocurrencies have many traits in common, they may vary in terms of their market cap, user base, application protocols, and mining rewards. In our dataset, we found two cryptocurrencies, namely Monero and JSEcoin, which are used for in-browser \cj. In \autoref{tab:cct}, we report the differences among the two cryptocurrencies. While both of them are used for \cj, 
at the time of writing of this paper, JSEcoin was not launched in the market and did not have any ``Initial Coin Offering'' (ICO), which explains its low prevalence in our dataset. Furthermore, unlike Monero, which is resource-intensive, JSEcoin uses minimal CPU power and does not add a significant processing overhead to the target device. One of the key objectives in this paper is to characterize resources abuse in \cc mining, where Monero is shown to be a better example than the ``browser-friendly'' JSEcoin. Therefore, due to its high prevalence in dataset, and the significant contribution towards the broader goal of this study, we mainly focus our work on Monero \cc. 


\begin{table}[t]
\small
\begin{center}
\caption{Detailed results of currency-based analysis.  {$^1$ The variable name is abbreviated. No CJ: No cryptojacking. }}
\label{tab:currencies}
\scalebox{0.9}{
\begin{tabular}{lrr|crr}
\toprule
\multirow{2}{*}{Platform} & \multicolumn{2}{c|} {Websites} & \multirow{2}{*}{\Cc} & \multicolumn{2}{c} {Websites} \\
                           & \# & \% & & \# & \% \\
                        
\midrule
Coinhive    & 4652 & 81.57 & \multirow{7}{*}{Monero} & \multirow{7}{*}{4926}  & \multirow{7}{*}{86.37} \\ 
Hashing     & 67 & 1.17 &                          &                          &   \\ 
deepMiner   & 56 & 0.98 &                          &                          &   \\ 
Freecontent & 39 & 0.68 &                          &                          &   \\ 
Cryptoloot  & 38 & 0.67 &                          &                          &   \\ 
Miner       & 38 & 0.67  &                         &                          &   \\ 
Authedmine  & 35 & 0.61 &                          &                          &   \\ \hline
JSEcoin     & 149 & 2.61  & JSEcoin                & 149                      &  2.61 \\ \hline
No CJ   & 628 & 11.01  &              ---      & 628                       & 11.01      \\ \hline
Total       & 5703 & 100.00  &              ---      & 5703                       & 100.00      \\ 
\bottomrule

\end{tabular}}
\end{center}
\vs{4}
\end{table}

\subsection{Code-based Analysis}
We perform static analysis on the \cj scripts to analyze the performance and complexity of their code. Static analysis reveals standard code-specific features that provide deeper insights into the flow of information upon code execution. For static analysis, we gathered \cj scripts from all the major \cj service providers found in our dataset, such as \ch, JSEcoin, Crypto-Loot, Hashing, deepMiner, Freecontent, Miner, and Authedmine. We observed that all the service providers had unique codes, specific to their own platform.  In other words, the websites using \ch{}'s services had the same \js code template across all of them. Therefore, $\approx$81.57\% of the websites in our dataset were using the same \js template for \cj. Similarly, all the websites using JSEcoin used the same standard template for their mining. However, the code template of each service provider was different from one another, which led us to believe that each script had unique static features. With all of that in mind, we performed static analysis on the \cj websites and compared the results with other standard \js for a baseline comparison.  

\begin{table}[t]

\caption{Comparison of Moneroe and JSEcoin. JSEcoin has not been released in the market as yet. }
\label{tab:cct}
\scalebox{0.9}{
\begin{tabular}{|l|l|l|c|l|}
\hline
\textbf{Currency} & \textbf{\begin{tabular}[c]{@{}l@{}}Market \\ Cap\end{tabular}} & \textbf{\begin{tabular}[c]{@{}l@{}}Consensus\\ Algorithm\end{tabular}} & \textbf{\begin{tabular}[c]{@{}l@{}}Resource\\ Intensive\end{tabular}} & \textbf{\begin{tabular}[c]{@{}l@{}}Dataset\\ Prevelance\end{tabular}} \\ \hline
Monero            & 2.3B     & CryptoNight  &   \cmark & 86.37\%  \\ \hline
JSEcoin           & ---     & SHA-256    & \xmark  & 2.61\%   \\ \hline
\end{tabular}}
\end{table}

\begin{table*}[htb]
\centering
\caption{Features extracted from cryptojacking, malicious, and benign samples for static analysis. The left most column shows the title of the features in each class while the remaining columns show the features extracted from \pl. Mean (($\mu$)) and Standard Deviation ($\sigma$) of the features are reported. The features obtained from these tables were used to perform correlation analysis and FCM clustering.}
\label{tab:fcj}
\scalebox{0.65}{
\begin{tabular}{|l|l|r|r|r|r|r|r|r|r|r|r|r|r|r|r|r|r|r|}
\hline
Cat.   & Platforms& $M$   & $M_d$   & $B$      & $D$       & $E$           & $c_l$   & $T$          & $\eta$   & $V$         &  $\eta _{1}$  & $n_1$   &  $\eta _{2}$ & $n_2$   & params & sloc & physical & $M_s$     \\ \hline

\multirow{10}{*}{\begin{turn}{-90} Cryptojacking \end{turn}} 
&deepMiner   & 184 & 44.2 & 14.1  & 113.0 & 4,810,434 & 4,667 & 267,246 & 554 & 42,533   & 47 & 2,440 & 507  & 2,227 & 75  & 416  & 499 & 67.8 \\ \cline{2-19}
&Authedmine  & 168 & 26.5 & 19.7  & 82.8  & 4,912,255 & 6,096 & 272,903   & 844 & 59,259  & 41 & 3,247   & 803 & 2,849  & 73  & 633  & 784 & 62.8 \\ \cline{2-19}
&Hashing     & 138 & 29.1 & 7.2   & 94.6  & 2,185,379 & 2,794 & 124,138 & 342 & 24,393   & 38 & 1,469 & 315  & 1,415 & 37  & 412  & 505 & 68.2 \\ \cline{2-19}
&Miner       & 133 & 27.7 & 9.3   & 90.5  & 2,537,930 & 3,239 & 140,996 & 403 & 28,032   & 39 & 1,690 & 364  & 1,549 & 49  & 479  & 617 & 64.1 \\ \cline{2-19}
&Coinhive    & 131 & 27.5 & 9.1   & 94.8  & 2,608,021   & 3,226 & 144,890    & 368 & 274,970      & 37 & 1,697   & 331 & 1,529  & 48  & 476  & 594 & 63.7 \\ \cline{2-19}
&Crypto-loot & 128 & 39.7 & 11.4  & 88.1  & 3,034,935 & 3,788 & 168,607 & 546 & 34,443   & 45 & 1,962 & 501  & 1,826 & 62  & 322  & 389 & 70.3 \\ \cline{2-19}
&Freecontent & 117 & 28.3 & 8.1   & 89.4  & 2,180,394 & 2,884 & 121,133   & 350 & 24,373   & 38 & 1,469 & 312  & 1,415 & 37  & 412  & 505 & 62.7 \\ \cline{2-19}
&JSEcoin     & 64  & 17.2 & 10.2 & 62.9 & 1,945,165   & 3,257 & 108,064 & 716 & 30,888   & 45 & 1,878 & 671  & 1,379 & 49  & 372  & 412 & 64.7 \\ \cline{2-19}
&\textbf{Mean} ($\mu$)	     &130.3	&	29.9	&	11.3	&	88.9	&	3,026,191	&	3,755.1	&	168,121	&	516.4	&	33,925	&	41.3	&	1,981.5	&	475.1	&	1,773.6	&	53.8	&	440.3	&	538.1	&	64.9 \\ \cline{2-19}
&\textbf{SD.} ($\sigma$)	&	35.9	&	8.4	&	3.9	&	13.8	&	1,180,403	&	1,109.9	&	65,577	&	185.1	&	11,856	&	3.9	&	599.3	&	182.8	&	519.3	&	14.8	&	93.2	&	126.3	&	2.8 \\
 \hline
\multirow{8}{*}{\begin{turn}{-90} Malicious \end{turn}} 
    & 20160209  & 92 & 21.5 & 5.6  & 25.1 & 423,925   & 1,833 & 23,551 & 580  & 16,826 & 27 &  1,032 & 553   & 801  & 22  & 427  & 503  & 44.4 \\ \cline{2-19}
    & 20161126  & 62 & 15.3 & 4.2  & 24.6 & 315,735   & 1,563 & 17,540 & 292  & 12,800 & 17 &  798 & 275     & 765  & 0   & 403  & 481  & 90.5 \\ \cline{2-19}
    & 20170110  & 14 & 4.4  & 15.0   & 26.7 & 1,211,305 & 4,704 & 67,294 & 782  & 45,210 & 15 &  2,740 & 767   & 1,964& 232 & 313  & 564  & 93.6 \\ \cline{2-19}
    & 20170507  & 6  & 24.0   & 5.9  & 11.1 & 199,917   & 1,864 & 11,106 & 777  & 17,897 & 18 &  942 & 759     & 922  & 1   & 25   & 890  & 71.7 \\ \cline{2-19}
    & 20160927  & 3  & 1.4  & 4.0   & 32.5 & 393,555   & 1,575 & 21,864 & 204  & 12,084   & 13 &  957 & 191     & 618  & 0   & 213  & 98   & 23.2 \\ \cline{2-19}
    & 20170322 & 2  & 18.1 & 11.8 & 7.1  & 253,442   & 3,514 & 14,080 & 1,123 & 35,607 & 9  &  1,762 & 1,114    & 1,752& 3   & 11   & 1,738& 90.9 \\ \cline{2-19}
    & 20170303  & 2  & 8.6  & 0.2  & 9.4  & 8,338     & 147   & 463    & 63   & 878    & 13 &   73 & 50      & 74   & 4   & 23   & 55   & 78.7 \\ \cline{2-19}
    & 20160407  & 1  & 33.3 & 0.1  & 2.7  & 207       & 19    & 11     & 16   & 76       & 5  &  12    & 11       & 7    & 0   & 3    & 3    & 78.9 \\ \cline{2-19}
    & 20170501  & 1  & 0.9  & 2.1  & 3.3  & 21,464     & 758   & 1,192   & 322  & 6,314   & 5  &    431 & 317      & 327  & 0   & 105  & 105  & 35.9 \\ \cline{2-19}
    & 20160810  & 1  & 12.5 & 0.5  & 11.9 & 20,148    & 275   & 1,119  & 70   & 1,685  & 6  &    255 & 64      & 20   & 0   & 8    & 13   & 60.4 \\ \cline{2-19}
&\textbf{Mean} ($\mu$)	&	18.4	&	14	&	4.9	&	15.5	&	284,803.7	&	1,625.2	&	15,822	&	422.9	&	14,938	&	12.8	&	900.2	&	410.1	&	725	&	26.2	&	153.1	&	445	&	66.9\\ \cline{2-19}
&\textbf{SD.} ($\sigma$)	&	31.9	&	10.5	&	5	&	10.8	&	364,470.8	&	1,508.9	&	20,248	&	374.8	&	15,045	&	6.9	&	834.7	&	372.5	&	686.6	&	72.6	&	171.9	&	543.5	&	24.9\\  
\hline
\multirow{8}{*}{\begin{turn}{-90} Benign \end{turn}} 
    & The Boat        & 2,135 & 69.3  & 110.8 & 392.0   & 130,285,522 & 31,916 & 7,238,084   & 1,364 & 332,361   & 59 &  17,341    & 1,305   & 14,575  & 852 & 3,084 & 3,349 & 66.7 \\ \cline{2-19}
    & IBM Design      & 2,119 & 68.3  & 110.9 & 397.1 & 132,237,213 & 32,018 & 7,346,511   & 1,351 & 332,981   & 59 &   17,393   &    1,292 & 1,4625 & 853 & 3,103 & 3,372 & 66.7 \\ \cline{2-19}
    & Histography     & 1,743  & 40.7  & 95.2  & 249.5 & 71,325,242    & 26,627  & 3,962,513     & 1,704 & 285,833    & 55 &  14,963  & 1,649    & 11,663     & 803 & 4,278  & 5,043  & 59.4 \\ \cline{2-19}
    & Know Lupus      & 1,006 & 28.1  & 92.9  & 170.4 & 47,474,425  & 25,120 & 2,637,468   & 2,181 & 278,600    & 54 &  13,424    & 2,127   & 11,696  & 615 & 3,583 & 4,288 & 65.2 \\ \cline{2-19}
    & tota11y         & 815   & 38.8  & 59.4  & 227.7 & 40,563,065  & 17,486 & 2,253,503   & 1,167 & 178,157   & 52 &   9,764   & 1,115    & 7,722  & 412 & 2,099 & 2,336 & 62.9 \\ \cline{2-19}
    & Masi Tupungato  & 784   & 58.2  & 47.1  & 185.0   & 26,199,193  & 14,296 & 1,455,510 & 958  & 141,585   & 43 &  7,875    & 915    & 6,421  & 238 & 1,347 & 1,470 & 67.2 \\ \cline{2-19}
    & Fillipo         & 703   & 42.9  & 43.1  & 194.3 & 25,139,766  & 12,900 & 1,396,653   & 1,045 & 129,377   & 54 &  7,132      &   991   & 5,768 & 269 & 1,637 & 1,770 & 61.5 \\ \cline{2-19}
    & Leg Work        & 412   & 75.7  & 34.0    & 241.3 & 24,651,056  & 11,100 & 1,369,503   & 589  & 102,143   & 45 &     5,835   & 544    & 5,265  & 66  & 544   & 633   & 65.9 \\ \cline{2-19}
    & Code Conf       & 409   & 27.8  & 41.1  & 197.1 & 24,336,420  & 12,500 & 1,352,023   & 939  & 123,437   & 49 &  7,162      &    890  & 5,338  & 315 & 1,469 & 1,753 & 64.9 \\ \cline{2-19}
    & Louis Browns    & 368   & 35.6  & 21.2  & 106.7 & 6,792,400   & 6,529  & 377,355     & 862  & 63,667    & 51 &   3,393     &    811  & 3,136  & 68  & 1,034 & 1,357 & 53.3 \\ \cline{2-19}
    & \textbf{Mean} ($\mu$)	&	1,049.4	&	48.5	&	65.6	&	236.1	&	52,900,430	&	19,049.2	&	2,938,912	&	1,216	&	196,814	&	52.1	&	10,428.2	&	1,163.9	&	8,621	&	449.1	&	2,217.8	&	2,537.1	&	63.4\\ \cline{2-19}
& \textbf{SD.} ($\sigma$)	&	694	&	17.8	&	33.6	&	92.8	&	44,755,377	&	9,151.2	&	2,486,409	&	459.8	&	100,856	&	5.3	&	4,999	&	456.7	&	4,165	&	310.3	&	1,225.4	&	1,418.2	&	4.3\\ 
\hline
\end{tabular}}
\end{table*}

\subsubsection{Data Attributes} \label{sec:dc}
We prepared our dataset for static analysis by collecting all of the popular \cj scripts from our list of websites. We found eight unique scripts among all the websites, each of which belongs to one of the service providers. As a control experiment, we collected an equal number of malicious and benign \js codes to design a clustering algorithm. Our aim was to obtain a set of features that were unique only to the \cj scripts, and aid in their detection. With such knowledge of those features, more accurate countermeasures can be further developed that will accurately predict if a given host machine is under \cj attack. 
To avoid bias towards a certain class, we were limited to include equal size of malicious and benign \js samples for the static analysis. Although there are many samples of malicious and benign \js in the wild~\cite{CurtsingerLZS11}, only eight \cj scripts are available in comparison. Since our work is focused on distinguishing \cj scripts from both malicious and benign \js, we had to balance the size of each class. While the number of scripts might seem as a limitation of our work, we believe the promise of this work is substantial: as more currencies and platforms start to use \cj, more samples will be available for a broader study. Demonstrating a baseline analysis to support the argument that \cj scripts are uniquely identifiable can open for further analysis of \cj scripts across well-understood analysis tools, which we explore in this paper.

In lieu, we used the existing data of the \cj websites (\textsection\ref{sec:data}) and online resources from GitHub for malicious \js sample \cite{Wizche17,Petrak17} . For benign \js, we used the set of non-cryptojacking websites and parsed their HTML code to extract benign \js code~\cite{staff_2017}. In summary, we had 8 samples of \cj~\js samples, spanning all the websites. Accordingly, we selected 10 malicious and 10 benign scripts for our clustering analysis (serves as a multi-class classification).

\subsubsection{Feature Extraction} \label{sec:feat}
We use various features that provide insights into the structure of the code and its maintainability. In the following, we describe the features we extracted for our static analysis of \cj, malicious, and benign scripts.

\BfPara{Cyclomatic Complexity} Cyclomatic complexity~\cite{McCabe76,WatsonMW96} measures the complexity of code  using Control Flow Gaph (CFG). It relies on a directed flow graph where each node represents a function to be executed and a directed edge between the two nodes indicates that the node representing the function will be executed after the previous node. Let $E$ be the number of edges, $N$ be the number of nodes, and $Q$ be the number of connected components in the CFG of a program, then $M$ can be used to denote the cyclomatic complexity of the program, and is calculated as $M = E + 2Q - N$.

\begin{figure*}[!t]
\hfill
\begin{subfigure}[Correlation in \cj JavaScript
\label{fig:crypto}]{\includegraphics[width=0.32\textwidth]{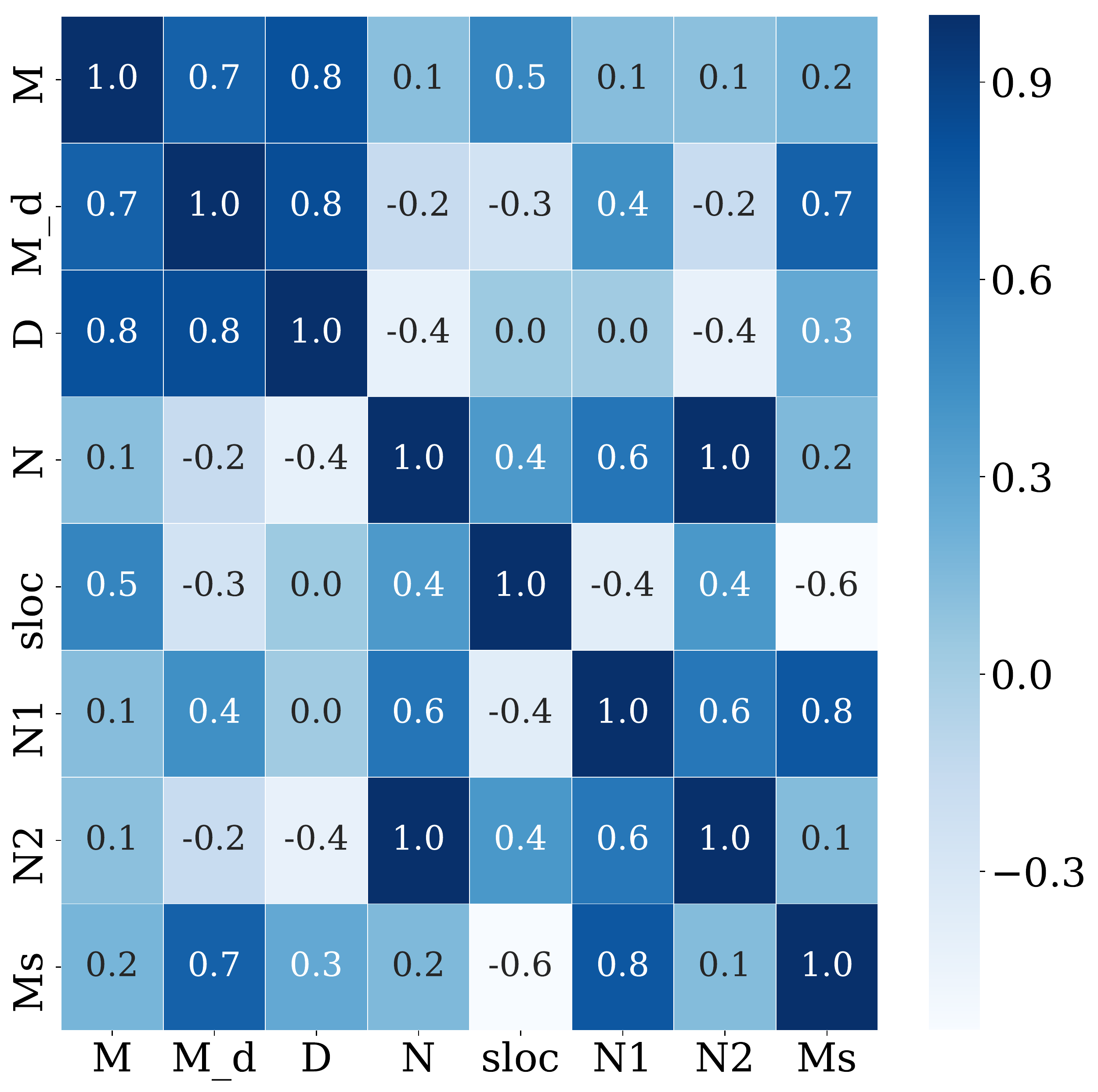}} 
\hfill
\end{subfigure}
\begin{subfigure}[Correlation in malicious JavaScript \label{fig:maliciousy}]{\includegraphics[width=0.32\textwidth]{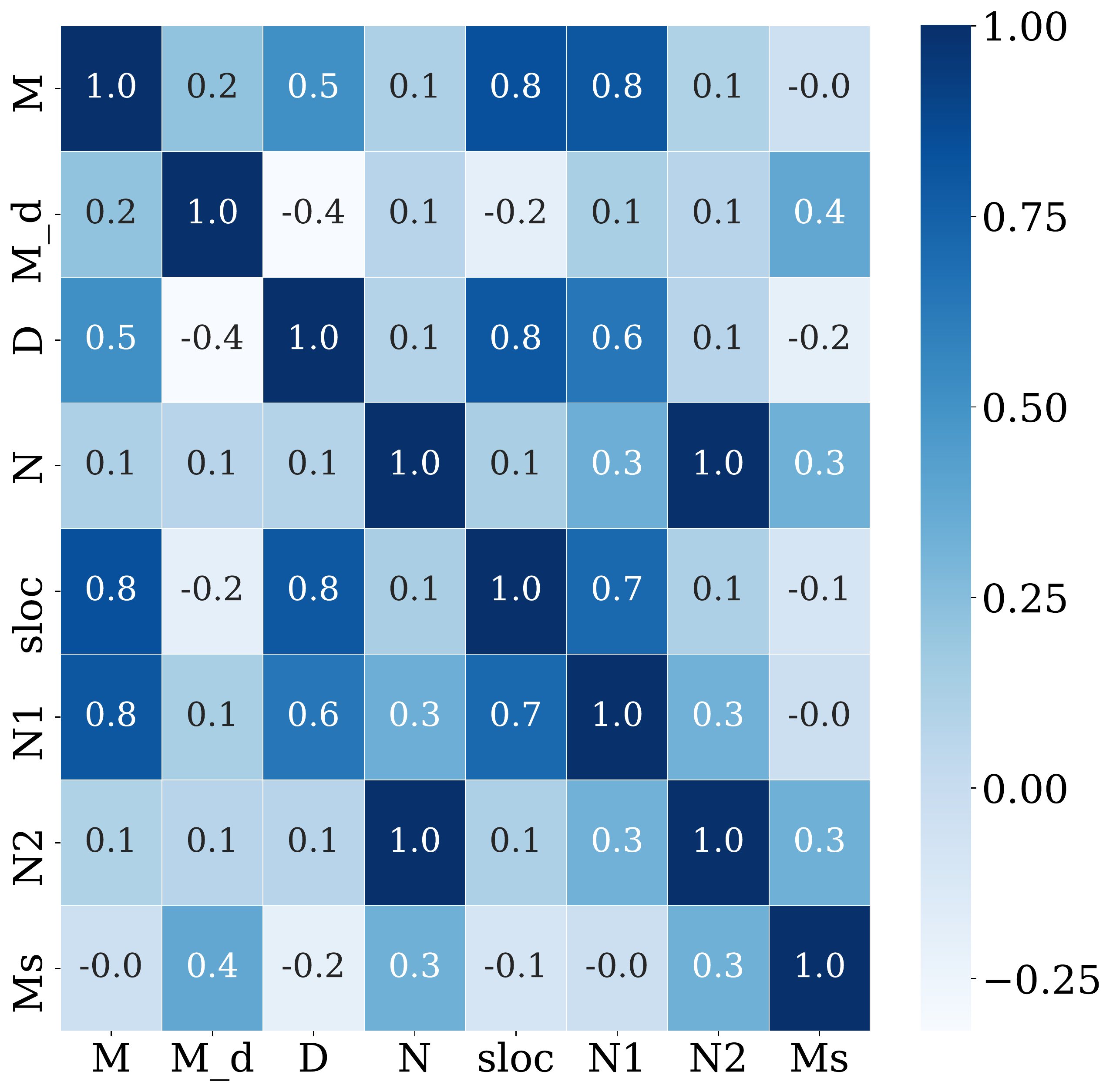}}
\hfill
\end{subfigure}
\begin{subfigure}[Correlation in benign JavaScript \label{fig:benign}]{\includegraphics[width=0.32\textwidth]{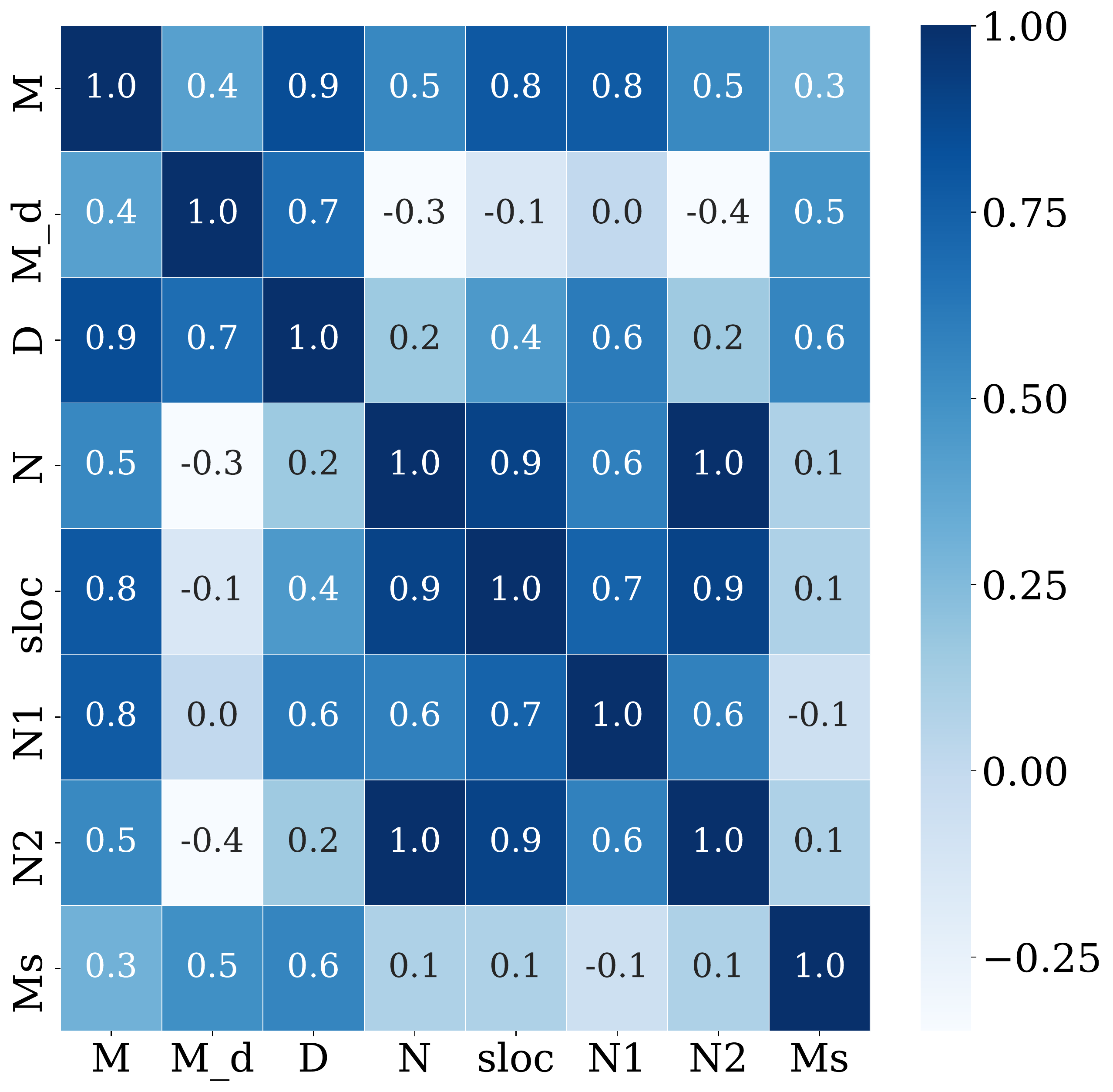}} 
\hfill
\end{subfigure}

\caption{Heatmap of correlation coefficients among the features of three categorizes of JavaScript. These are the subset features of \autoref{tab:fcj}, obtained by using \autoref{algo:corr}. It can be noted that features among benign scripts appear to be highly correlated while the features among malicious scripts remain highly uncorrelated. Correlation among the features of cryptojacking scripts remains in the middle, relative to the other two.    } 
\label{fig:correlation}

\end{figure*}

\BfPara{Cyclomatic Complexity Density} Cyclomatic complexity density~\cite{FentonN99} is a measure of Cyclomatic complexity, defined above, spread over the total code length. Usually, malware authors obfuscate their code to avoid detection. As such, among many other possibilities of obfuscation, they may alter the flow of a program and add extra functions. While adding more functions and lines of code will certainly increase the size of the code, its complexity will remain the same, which could be used as a feature of their detection. Let $c_l$ be the total number of lines of code, then the cyclomatic complexity density, denoted by $M_d$, can be computed as $M_d = \frac{E + 2P - N}{c_l}$

\BfPara{Halstead Complexity Measures} In software testing, the Halstead complexity measures are used as metrics to characterize the algorithmic implementation of a programming language~\cite{Serebrenik11}. Those measures include the vocabulary $\eta$, the program length $n$, the calculated program length $n_c$, the volume $V$, the effort $E$, the delivered bugs $B$, the time $T$, and the difficulty $D$. Let the number of distinct operators be $\eta _{1}$, the number of distinct operands be $\eta _{2}$, the total number of operators be $n_1$, the total number of operands be $n_2$, the $\eta, n, n_l, V, E,$ and $B$ are defined as follows: 
\begin{align}\label{equation:halsted} 
    \nonumber\eta &= \eta _{1}+ \eta _{2},  & n &= n_{1}+n_{2}\\
    \nonumber n_c &= (\eta _{1}\log _{2}\eta _{1})+(\eta _{2}\log _{2}\eta _{2}), & V &= n\times \log _{2}\eta\\
    \nonumber D &= (\eta _{1}/ 2)\times (n_{2} / \eta _{2}), & E &= D\times V\\
    \nonumber T &= {(D \times V)}/{18}, & B &= {E^{\frac{2}{3}}}/{3000} 
\end{align}

\BfPara{Maintainability Score} The maintainability score $M_s$ is calculated using Halstead volume $V$, cyclomatic complexity $M$, and the total lines of code in the \js file $c_l$. The maintainability score index $M_i$ is calculated between [0-100] and is defined as: 
\begin{align} 
\nonumber M_s & =  171 - 5.2 \log (V) - 0.23M  - 16.2 \log (c_l);
\\ \nonumber M_i &= \max(0, \frac{M_s}{171});
\end{align}

\BfPara{Source Lines of Code} Source lines of code (SLOC) is a measure of the lines of code in the program after excluding the white spaces. SLOC is used as a predictive parameter to evaluate the effort required to execute the program. It also provides insights about program maintainability and productivity.

\subsubsection*{Results} To extract the aforementioned features in our code-based analysis, we used \pl, a \js static analysis and source code complexity tool~\cite{badge_2016}. For each \js code, we run \pl and record the 17 extracted features, highlighted above, as reported in~\autoref{tab:fcj}. From~\autoref{tab:fcj}, we observed that certain features, such as $M$, $M_d$, $V$, and $T$, are clearly discriminative among all the categories. For further analysis, in the next section we will look into the correlation of these features among each category to see whether there is a unique pattern among each category, which allow us to build a clustering system that can automatically identify different \js categories based on the extracted features.

\subsubsection{Correlation Analysis} \label{sec:correlation}
Presenting individual features among those analyzed above, while meaningful, might not shed light on their distinguishing power given their large numbers. To this end, we  pursue a correlation analysis to understand their patterns. In particular, we conducted a correlation analysis to observe the similarity of features among the three categorizes of scripts, the \cj, malicious, and benign. The correlation analysis showed the consistency of the relationship distinctive to each category of the \js codes. As such, this provided us with insights into coding patterns and features unique to the style of coding \cj scripts, malware scripts, and benign scripts. We computed the correlation of the features in all the scripts belonging to each category of \js. We used the Pearson correlation coefficient for this analysis, which is defined as $\rho(X,Y) = {{Cov}(X,Y)}/({\sqrt{{Var}(X){Var}(Y)}})$, where $X$ and $Y$ are the random variables, $Var$ and $Cov$ are the {\em variance} and {\em covariance} of the random variables, respectively. 

To identify distinguishing features and reason about their prevalence in \cj~\js, we performed comparative analysis on the correlation matrix obtained for each class. In \autoref{algo:corr}, we outline the procedure used to identify those features. The algorithm takes as an input the correlation matrix of \cj \textbf{C}, malicious \textbf{M}, and benign \textbf{B} \js features reported in \autoref{tab:fcj}, computes the mean of the column vector with respect to one feature in the row, compares the mean feature of each class, and outputs the most distinguishing features in \cj scripts that are highly correlated within their class. The distinguishing aspect of a feature in cryptojacking class is obtained by subtracting its mean value from complementary mean values of features from the other two classes, and selecting the maximum difference.

\begin{algorithm}[t]  

\SetKwInOut{Input}{Inputs}  
\Input{\textbf{C}, \textbf{M}, \textbf{B};\\
}
$i = len(\textbf{C}$)\;

$Cmean,Mmean,Bmean,Array$ = []\;

\For{$k = 0;\ k < i;\ k = k + 1$}{
$Cmean [k] \leftarrow \dfrac{(\sum_{j=1}^{{i}} {c(i,j)}} {{i}}$\;
$Mmean [k]\leftarrow \dfrac{(\sum_{j=1}^{{i}} {m(i,j)}} {{i}}$\;
$Bmean [k]\leftarrow\dfrac{(\sum_{j=1}^{{i}} {c(i,j)}} {{i}}$\;
\If{$(Cmean [k] - Mmean [k] \&\& Cmean [k] - Bmean [k]) >  (Mmean [k] - Bmean [k])$ }{
$Array \leftarrow Cmean [k]$\;}}
\SetKwInput{KwData}{Output}
 \KwData{$Array$ }

  \caption{Identifying Significant Features}
\label{algo:corr}
\end{algorithm}

The output $Array$, in \autoref{algo:corr}, contains a subset of features from the total seventeen features that are unique to \cj scripts. In particular, we found eight features and plotted their result in~\autoref{fig:correlation}. It can be observed that \cj scripts are more correlated with respect to the cyclomatic complexity density $M_d$ and the maintainability score $M_s$, while malicious and benign scripts are not as correlated over those same parameters. From the description of those features provided in \textsection\ref{sec:feat}, deeper insights can be developed regarding the coding patterns, code complexity, CFGs, and maintainability of \cj scripts. Furthermore, high correlation also provides a valuable insight into code contents: that all \cj scripts must be performing a sequence of similar actions with complementary execution patterns and information flows. We apply this understanding in our dynamic analysis and validate it using WebSocket inspection.

\begin{table}[t]
\small
\begin{center}
\caption{Confusion matrix and evaluation metrics of the cryptojacking (CJ), malicious, and benign scripts' clustering results based on FCM clustering algorithm. Evaluation metrics' names are abbreviated. FPR= False Positive Rate, FNR= False Negative Rate, and AR=Accuracy Rate.}
\scalebox{0.93}{
\begin{tabular}{l|ccc|ccc}
\toprule
	Class	&	Benign	&	Malicious	&	CJ &	FPR	&	FNR	&	AR	\\ \hline

Benign  	&	9	&	0	&	1	&	10	&	0	&	90	\\ 
Malicious	&	0	&	10	&	0	&	0	&	0	&	100	\\ 
CJ	    &	0	&	0	&	8	&	0	&	11.1	&	100	\\ \cline{1-7}
Total   	&		\multicolumn{3}{c|}{}		&	3.3	&	3.7	&	96.42	\\ 
\bottomrule
\end{tabular}}
\label{tab:confusionmatrix}  
\end{center}
\end{table}

\subsection{Clustering} \label{sec:clustering}
In this section, we build a classification system that automatically recognizes \cj scripts from malicious and benign scripts based on the code complexity features alone, which could be easily extracted from the \cj scripts and are common among a large number of \cj websites. It is desirable for our classification system to classify scripts even  with minimal information regarding the labels of the scripts. Therefore, we  utilized the Fuzzy C-Means (FCM) clustering algorithm~\cite{bezdekEF84}, which has the advantage of being an unsupervised learning algorithm. In the other words, in comparison with supervised classification algorithms, such as the Support Vector Machine (SVM) and Random Forest (RF), which require labels of the dataset in the training phase, FCM has the advantage of performing well on the unlabeled dataset.

The main goal of the FCM is to group a dataset $X$ into $C$ clusters in which every data point belongs to every cluster to a certain degree. In other words, a data point that lies close to the center of a certain cluster will have a higher membership degree to that cluster, whereas the membership degree of the data point that lies far away from the center of this cluster will be lower~\cite{bezdekEF84}. We utilized the FCM clustering algorithm to group the scripts to \cj, malicious, and benign clusters. In order to evaluate the performance of the clustering experiment, we used standard evaluation metrics; the confusion matrix, Accuracy Rate (AR), False Positive Rate (FPR), and False Negative Rate (FNR), which are reported in~\autoref{tab:confusionmatrix}. 

\begin{figure}[t]
\centering
\includegraphics[width=0.4\textwidth]{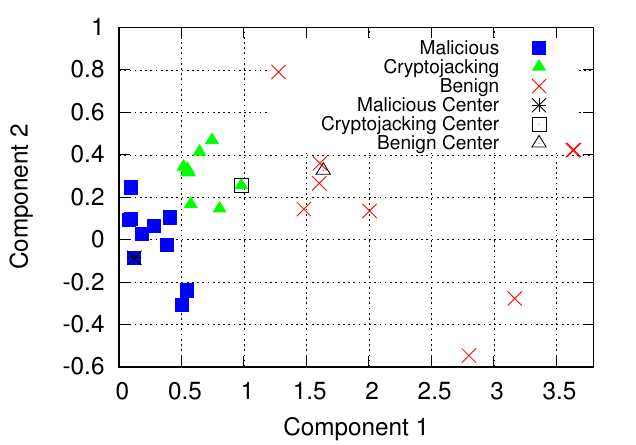}

\caption{Clustering of the cryptojacking, malicious, and benign scripts using FCM clustering algorithm.}
\vs{5}
\label{fig:clustering}   
\end{figure}

As shown in~\autoref{tab:confusionmatrix}, the clustering algorithm is able to identify the scripts with high performance: AR of $\approx$96.4\%, FPR of 3.3\%, and FNR of 3.7\%. In addition, we have visualized these clusters based on two major principal components of their features which, in~\autoref{fig:clustering}, clearly show natural separation between the clusters using the underlying features.

\section{Dynamic Analysis} \label{sec:dynamic}
Despite the clear benefits of the static analysis outlined above, it is limited, and subject to circumvention through \js code obfuscation. To this end, we conduct dynamic analysis that looks into profiling the usage of \cj~\js code of various host resources: CPU, and battery. We then look into the characteristics of \cj in their use of network resources.


\subsection{Resource Consumption Profiling} \label{sec:processor}

We conduct an extensive analysis of CPU and battery usage of the various \cj scripts.  

\subsubsection{Settings and Measurements Environment} 
We noticed that in each \cj website, a \js snippet encodes a key belonging to the code owner and a link to a server to  which the PoW is ultimately sent. \autoref{lst:coinhive} provides a script found in websites that use \ch for mining. The source (\textit{src}) refers to the actual \js file that is executed after a browser loads the script tag. In this script, we also noticed a {\em throttling parameter}, which is used as a mean of controlling how much resources a \cj script uses on the host. We use such a throttling parameter, $\alpha$ as an additional variable in our experiment. We experiment with $\alpha=\{0.1, 0.5, 0.9\}$.

To understand the impact of \cj on resources usage in different platforms, we use battery-powered machines running Microsoft Windows, Linux, and Android operating systems (OSes). For our experiments, we selected three laptops, each with one of those OSes. The Windows laptop used in the experiment was Asus V502U, with Intel Core i7-6500U processor operating at 3.16 GHz. The Linux laptop was Lenovo G50, with Intel Core i5-5200U processor (4 cores) running at 2.20 GHz, and the Android phone was {\em Samsung Galaxy J5}, with Android version of 6.0.1.

For our \cj script construction, using the various parameters learned above, we set up an account on \ch to obtain a key that links our ``experiment website'' to the server. Next, we set up a test website and embedded the code in~\autoref{lst:coinhive} within the HTML tags of the website. Finally, to measure the usage of resources while running \cj websites, we set up a Selenium-based web browser automation and run \cj websites, for various evaluations. Selenium is a portable web-testing software that mimics actual web browsers~\cite{bruns2009web,seleniumdocumentation}.

\subsubsection{CPU Usage} 
First, we baseline our study to highlight CPU usage as a fingerprint across multiple websites that employ \cj using the aforementioned configurations and measurement environment. We study the usage of CPU with and without \cj in place. For this experiment, we select four \cj websites. To measure the impact of \cj on CPU usage, we ran those websites in our \slx environment, for 30 seconds, with \js enabled (thus running the \cj scripts) and disabled (baseline; not running the \cj scripts). We use this test experiment as our control. 

\begin{figure}[t]
\begin{lstlisting} [caption={\ch code found in \cj sites.}, label={lst:coinhive},style=htmlcssjs,]
<script src="./Welcome_files/coinhive.min.js"></script>
<script>
	var miner = new coinhive.Anonymous("owner key", 
	    {throttle: 0.1});
		miner.start();
</script>
\end{lstlisting}
\end{figure}

\begin{figure}[t]
\begin{subfigure}[JavaScript enabled\label{fig:dyn}]{\includegraphics[width=0.23\textwidth]{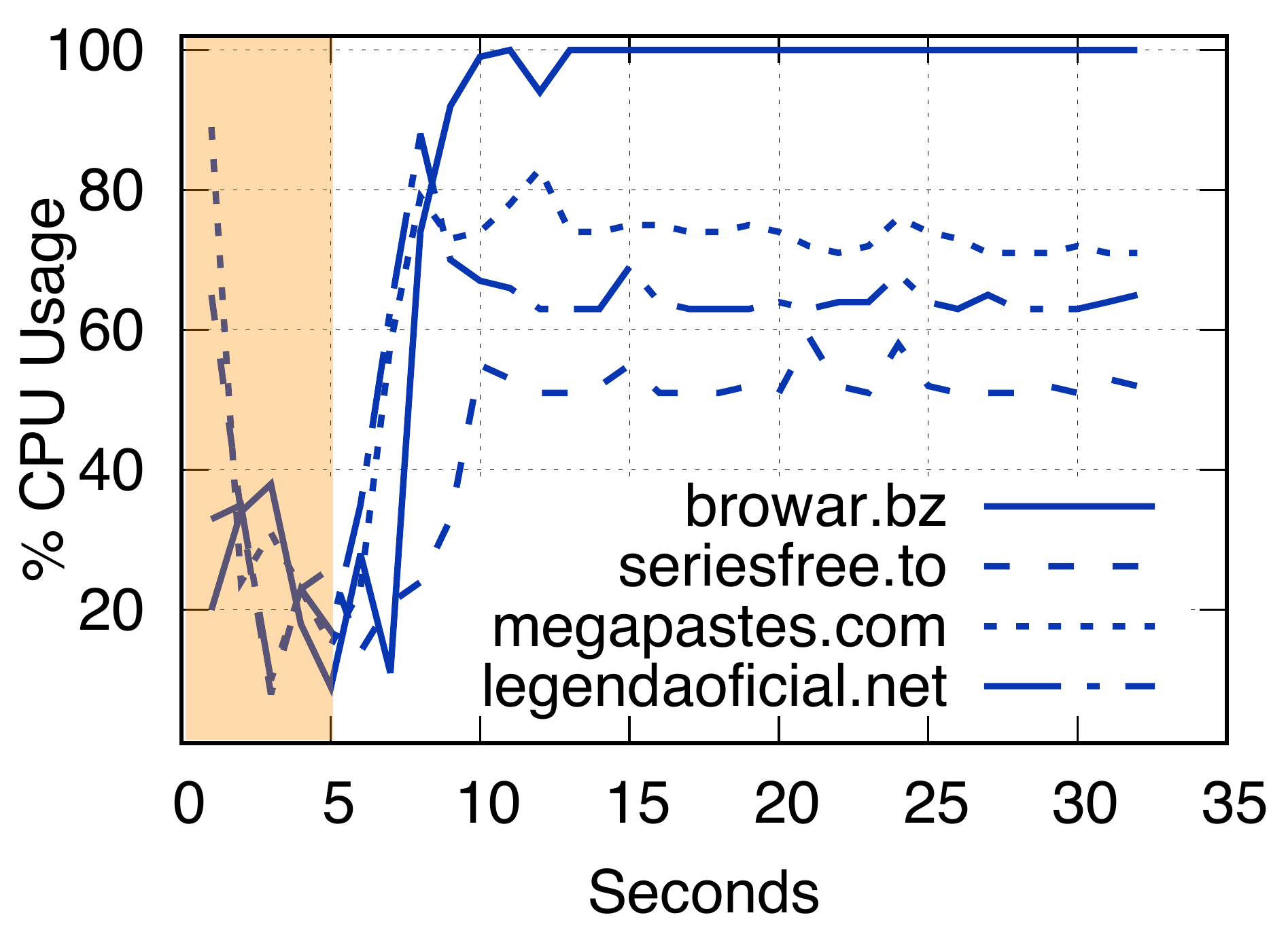}} 
\end{subfigure}
\begin{subfigure}[JavaScript disabled\label{fig:maliciousx}]{\includegraphics[width=0.23\textwidth]{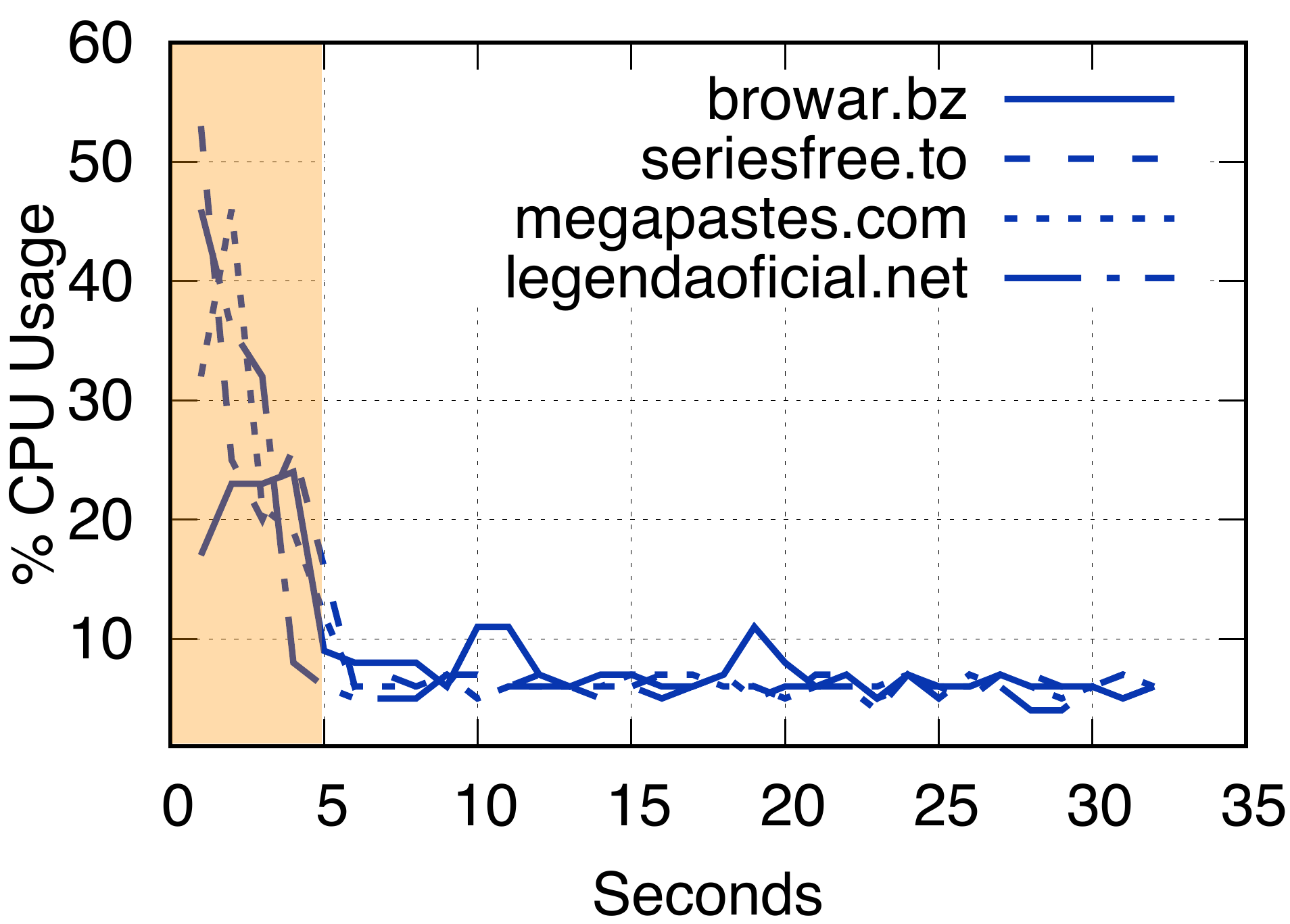}}
\end{subfigure}

\caption{Processor usage by four different cryptojacking websites with JavaScript enabled and disabled. Notice that with JavaScript enabled, the processor usage increases 10-20 times.}
\vs{3}
\label{fig:dynamic}
\end{figure}

\begin{figure*}[]
\centering
    	\subfigure[CPU usage on Windows \label{fig:windowscpu}] {\includegraphics[width=0.3\textwidth]{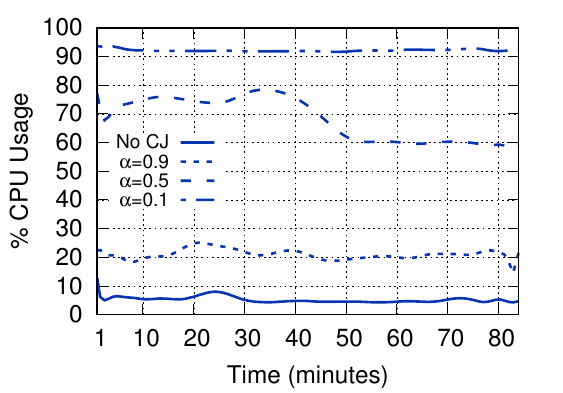}}
        \subfigure[CPU usage on Linux \label{fig:linuxcpu}] {\includegraphics[width=0.3\textwidth]{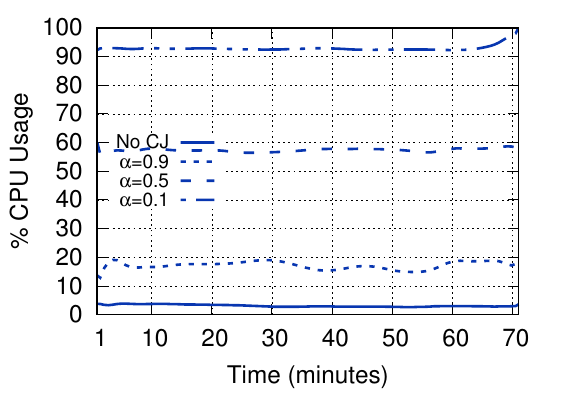}}
      	\subfigure[CPU usage on Android \label{fig:maccpu}] {\includegraphics[width=0.3\textwidth]{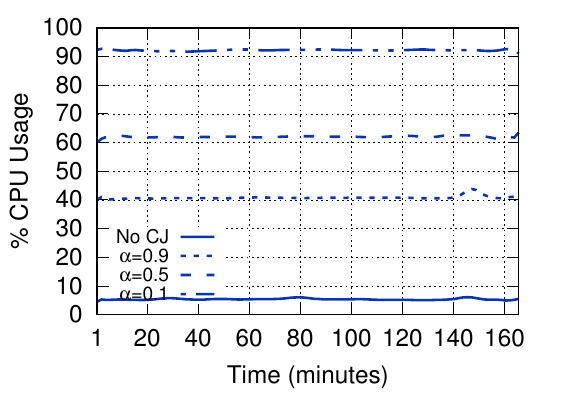}}

\caption{CPU usage recorded on three devices. Windows machine consumed more processing than the other devices during cryptojacking. This also explains the high battery drainage in~\autoref{fig:windowsbattery}. }
\vs{6}
\label{expriment:cpu}
\end{figure*}

\BfPara{Results} We obtained two sets of results for each website, with and without \cj. In~\autoref{fig:dynamic}, we plot four test samples obtained from our experiment to demonstrate the behavior of websites with and without \cj. From those results, we observe that when a website is loaded initially it consumes a significant CPU power (shaded region), in both cases. Once the website is loaded, the CPU consumption decays if the \js is disabled, indicating no \cj. When \js is enabled, the CPU consumption is high, indicating \cj. It can also be observed in~\autoref{fig:dynamic}, that the CPU usage varied across the websites, indicating the usage of the throttling parameter highlighted above. The same behavior as with \js disabled is exhibited when loading a page with \js that is either benign or of other types of maliciousness than \cj. Through this experiment, we found that \cj consumes anywhere between 10 and 20 times the processing power compared to when not using \cj on the same host. To further understand the impact of throttling on CPU usage in different platforms, we conduct another measurement where we used $\alpha=\{0.1, 0.5, 0.9\}$ with the different testing machines. We found a consistent pattern, whereby the relationship between $\alpha$ and the CPU usage is linear, as demonstrated in~\autoref{expriment:cpu}.

\subsubsection{Battery Usage} Clearly, high CPU usage translates to higher power consumption, and quicker battery drainage. To further investigate how \cj affects battery drainage, we carried out several experiments using various $\alpha$ values for the various platforms. Here we are interested in the order of battery drainage from a baseline, rather than comparing various platforms. The batteries of the different machines are as follows: 65 watt-hour for Windows, 41 watt-hour for Linux and $\approx$9.88\% watt-hour for Android. 

\begin{figure*}[]
\centering
		\subfigure[Battery consumption on Windows\label{fig:windowsbattery}] {\includegraphics[width=0.3\textwidth]{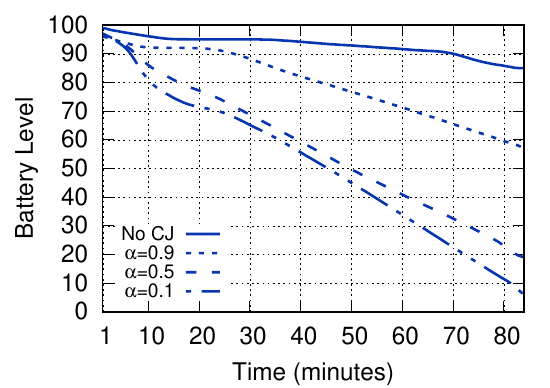}}
    	\subfigure[Battery consumption on Linux \label{fig:linuxbattery}] {\includegraphics[width=0.3\textwidth]{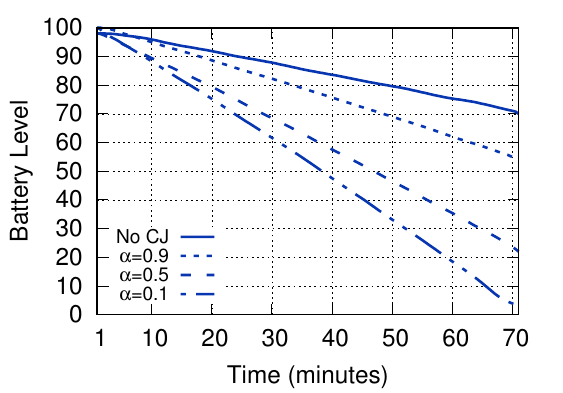}}
    	\subfigure[Battery consumption on Android \label{fig:macbattery}] {\includegraphics[width=0.3\textwidth]{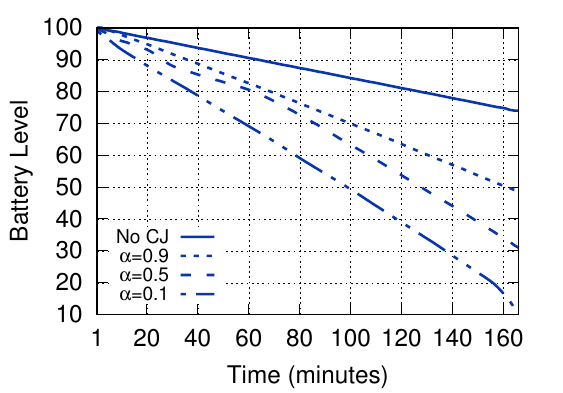}}

\caption{Battery usage recorded on three devices used in the dynamic analysis  Notice that Windows OS is mostly affected by \cj as its aggregate battery drainage is maximum compared to Linux and Mac. }
\label{expriment:battery}
\vs{6}
\end{figure*}

\BfPara{Results} For each $\alpha\in \{0.1, 0.5, 0.9\}$, and using the different devices, we ran the \js script on a fully charged battery. We logged the battery level every 30 seconds, as the script ran on each device with the given $\alpha$ value, starting from a fully-charged battery. Finally, we measure the baseline by running our script without the \cj code. The results are shown in~\autoref{expriment:battery}. As expected, with $\alpha=0.1$, corresponding to the lowest throttling and highest CPU usage, the battery drained very quickly, to $\approx$10\% of its capacity within 80 minutes, compared to $\approx$85\% within the same time when not using \cj. The same result is demonstrated for both the Linux laptop and Android phone. We also notice that relationship between $\alpha$ and the battery drainage is also linear.

In examining the CPU and battery usage by \cj websites, as shown above, we highlight a clear and unique patterns that can be used to identify those websites. We also notice that the different operating systems do not have any architectural  support to prevent activities like \cj from happening on the device. 

\subsection{Network Usage and Profiling} \label{sec:throt}
Dynamic network-based artifacts are essential in analyzing \cj scripts, especially when those scripts are obfuscated. To this end, we also explore the network-level artifacts to reconstruct the operation of \cj services.

We noticed that during \cj website execution, the \js code establishes a WebSocket connection with a remote server and preforms a bidirectional data transfer. The WebSocket communication can be monitored using traffic analyzers such as {\em Wireshark}. However, a major issue when using traffic analyzers is that browsers encrypt the web traffic during WebSocket communication. Although significant information can still be gathered, such as source, destination, payload size, and request timings, the actual data transferred remain encrypted, preventing further analysis. To perform a deeper analysis on WebSocket traffic, we examined the actual data frames {\em in the browser} to understand the communication protocol and payload content of WebSocket connection, for possible analysis of \cj websites, outlined below.

\begin{figure}[t]
\begin{lstlisting}[caption={WebSocket frames exchanged between the client and the server},label={lst:auth}, style=json]%[t]

// auth request from client to server 
{"type": "auth",
    "params": {
    "site_key": "32 characters key",
    "type": "anonymous", "user": null, "goal": 0 }}
// authed response from server to client    
{ "type": "authed",
	"params": {
	"token": "", "hashes": 0 }}
// job request sent by the server to client	
{ "type": "job",
	"params": {
	"job_id": "164698158344253",
	"blob": "152 characters blob string",
	"target": "ffffff00" }}
// submit message by client to server
{ "type": "submit",
	"params": {
	"job_id": "164698158344253", "nonce": "cfe539d3",
	"result": "256-bit hash" }}
// hash_accept sent by server to client
{ "type": "hash_accept",
	"params": {
	"hashes": 256 }}

\end{lstlisting}
\vs{8}
\end{figure}

When a WebSocket request is initiated, the client sends an {\em auth} message to the server along with the user information, including {\em sitekey}, {\em type}, and {\em user}. The length of {\em auth} message is 112 bytes. The {\em sitekey} parameter is used by the server to identify the actual user who owns the key of the \js and adds balance of hashes to the user's account. The server then authenticates the request parameters and responds back with {\em authed} message. The {\em authed} message length is 50 Bytes and it includes a token and the total number of hashes received so far from the client's machine. In the {\em authed} message,  the total number of hashes is 0, since the client has not sent any hashes yet. Then, the server sends {\em job} message to the client. The {\em job} message has a length of 234 Bytes with a {\em job\_id}, {\em blob}, and {\em target}. The {\em target} is a function of the current difficulty in the cryptocurrency to be mined. The client then computes hashes on the {\em nonce} and sends a {\em submit} message back to the server, with {\em job\_id}, {\em nonce}, and the resulting hash. The {\em submit} message has a payload length of 156 Bytes. In response to the {\em submit} message, the server sends {\em hash\_accept} message with an acknowledgement and the total number of hashes received during the session. The {\em hash\_accept} message is 48 Bytes long. This is to be noted that once a webpage is refreshed, the WebSocket connection terminated and restarted. On the other hand, if multiple tabs are opened in the same browser, the WebSocket connection remains unaffected. In~\autoref{tab:websocket}, we provide details about the WebSocket connection during a \cj session. In~\autoref{lst:auth}, we provide the the actual data frames exchanged between the browser and the server during WebSocket session. The data frames are structured in ``JavaScript Object Notation'' (JSON).  

\begin{table}[t]
\centering
\caption{Types of messages exchanged between the client and the server during cryptojacking WebSocket connection. Length is measured in byte.}
\label{tab:websocket}
\scalebox{0.90}{
\begin{tabular}{|l|l|l|c|l|}
\hline
\textbf{Message} & \textbf{Source} & \textbf{Sink} & \textbf{Length} & \textbf{Parameters}\\ \hline
{auth}      & client            & server        & 112                   & sitekey, type, user     \\ \hline
{authed}    & server          & client          & 50                    & token, hashes \\ \hline
{job}       & server          & client          & 234                   & job\_id, blob, target   \\ \hline
{submit}    & client            & server        & 156                   & job\_id, result  \\ \hline
{hash\_accept} & server          & client          & 48                    & hashes   \\ \hline
\end{tabular}}
\vs{3}
\end{table}

\begin{table*}[t]
\centering
\caption{Results of cryptojacking with different devices. {Here $\alpha$ is the throttling parameter, $h$, $\Delta{t}$, $b_n$, $b_c$, $W$, $P$, and $L$ are the parameters obtained from~\autoref{equ:profit} and~\autoref{equ:loss}. $T$ is the estimated time required for each device to mine 1 XMR}. }
\label{tab:eco}
\scalebox{0.85}{
\begin{tabular}{|c|c|c|c|c|c|c|c|c|c|c|}
\hline
\textbf{Device}  & $\Delta{t}$ (mins)  & $b_n (\%)$ & $\alpha$ & $h$ (hps) &  $b_c$ (\%) & $W$(W/h) & $P $(USD) & $L$ (USD) & $L-P$ (USD) & $T$(years)   \\ \hline
    
\multirow{3}{*}{\bf{Windows}}      & \multirow{3}{*}{85}           & \multirow{3}{*}{82}        & 0.1         & 21                  & 10          & 65          & 6.4 $\times 10^{-4}$    & 4.5 $\times10^{-3}$                             & 3.8 $\times10^{-3}$        & 50  \\ \cline{4-11} 
                                   &           &         & 0.5        & 14                   & 19          & 65          & 3.1 $\times 10^{-4}$    & 3.7 $\times10^{-3}$                                                                   & 3.4 $\times10^{-3}$         & 104                                        \\  \cline{4-11}
                                   &           &         & 0.9        & 5                    & 57          & 65          & 4.4 $\times 10^{-5}$    & 1.6 $\times10^{-3}$                                                                   & 1.5 $\times10^{-3}$         & 367                                 \\  \cline{1-11}

\multirow{3}{*}{\bf{Linux}}        & \multirow{3}{*}{71}          & \multirow{3}{*}{70}       & 0.1        & 26                   & 3           & 41          & 6.6 $\times 10^{-4}$    & 5.5 $\times10^{-3}$                               & 4.8 $\times10^{-3}$        & 40 \\ \cline{4-11}
                                   &           &         & 0.5        & 16                   & 22          & 41          & 4.1 $\times 10^{-4}$    & 4.2 $\times10^{-3}$                                                                    & 3.8 $\times10^{-3}$     & 66  \\ \cline{4-11}
                                   &           &         & 0.9        & 5                    & 54          & 41          & 1.3 $\times 10^{-4}$    & 2.6 $\times10^{-3}$                                                                   & 2.5 $\times10^{-3}$        & 214 \\ \cline{1-11}

\multirow{3}{*}{\bf{Android}}        & \multirow{3}{*}{163}          & \multirow{3}{*}{76}       & 0.1        &  5                  & 11           & 9.9          & 2.8 $\times 10^{-4}$    & 9.5 $\times10^{-4}$                              & 6.7 $\times10^{-4}$       & 220  \\ \cline{4-11}
                                     &           &         & 0.5        & 3                   & 32          & 9.9                                                & 1.7 $\times 10^{-4}$    & 7.2 $\times10^{-4}$                              & 5.5 $\times10^{-4}$      & 369      \\ \cline{4-11}
                                     &           &         & 0.9        & 2                    & 49          & 9.9                                               & 1.1 $\times 10^{-4}$    & 5.4 $\times10^{-4}$                              & 4.3 $\times10^{-4}$   & 574        \\ \cline{1-11}

\end{tabular}}
\vs{2}
\end{table*}

\vs{3}
\section{Economics of Cryptojacking} \label{sec:economic}
In this section, we evaluate the economic feasibility of \cj by extrapolating on the results in our dynamic analysis. We look at the economic feasibility from the perspective of a \cj website's owner, intentional \cj, malicious \cj, and website visitors. For \cj, the reward of the website owner or adversary depends on the number of hashes produced while a website visitor visits the website.  We formulate the analysis as a feasibility: how much of the energy consumed by \cj scripts (cost) is transferred to the \cj website owner, whether malicious or benign, and how that translates as an alternative to online advertisement. 

\vs{2}
\subsection{Analytical Model}\label{sec:ana}
\begin{figure}[t]
\begin{center}
\includegraphics[ width=0.35\textwidth]{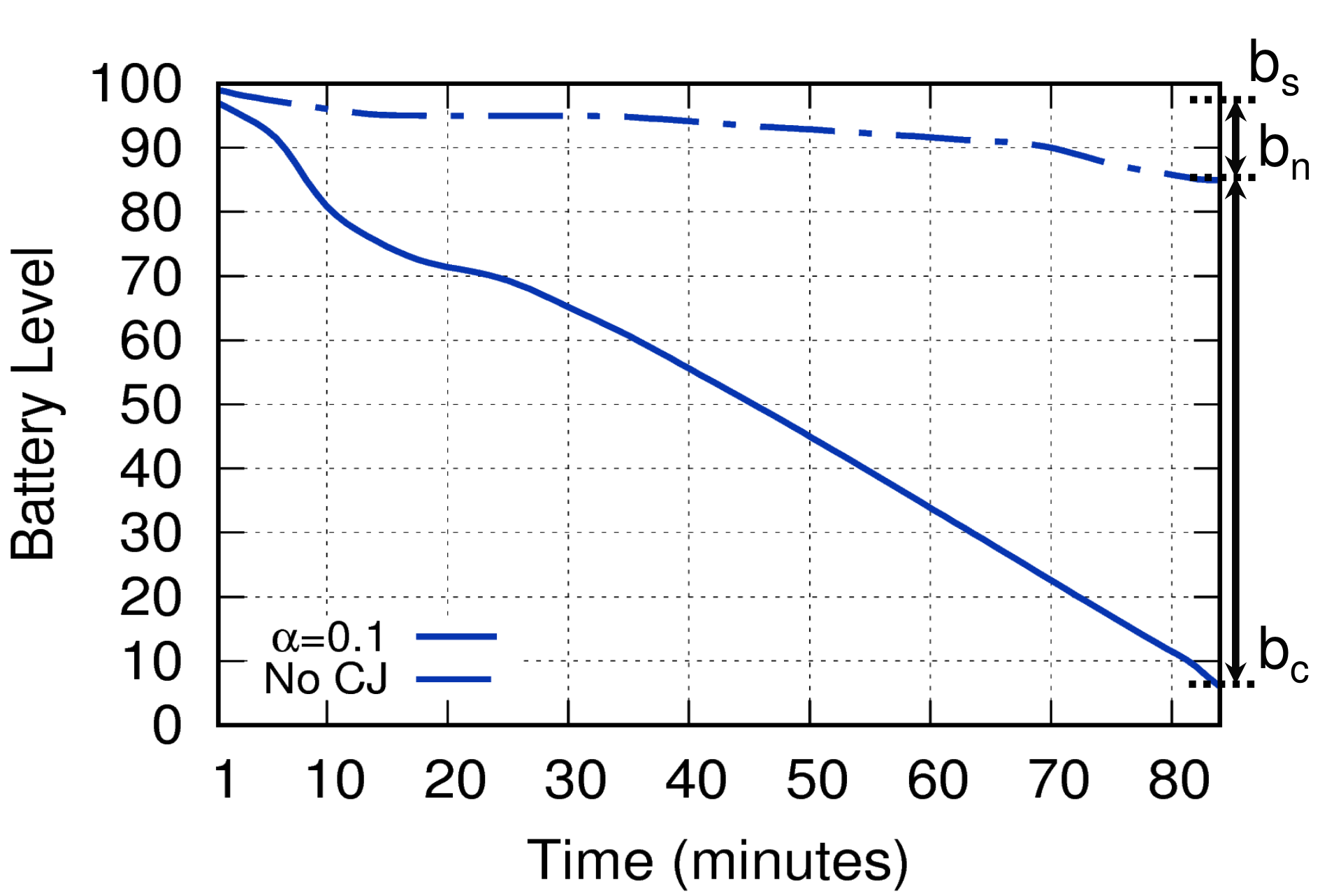}

\caption{Battery drain sample of Windows i7. {Here $b_s$ denotes the starting point of the battery, $b_n$ denotes the normal 80 minutes battery drain without cryptojacking and $b_c$ denotes the battery drain with maximum cryptojacking }.}
\vs{10}
\label{fig:overviewbattery}
\end{center}
\end{figure}

To set a stage for our analysis, in~\autoref{fig:overviewbattery} we present the results from one sample experiment conducted on Windows i7 machine with \cj website set to minimum throttling ($\alpha$=0.1), indicating a maximum \cj. In this figure, the region between $b_s$ and $b_n$ is a baseline, unrelated to \cj--due to normal operation of the system. On the other hand, the region between $b_n$ and $b_c$ is the battery drainage due to \cj. We refer to the energy loss due to such \cj as $L$ for a given user. To formulate the cost (to users) and benefit (to \cj website), let $P$ be the benefit (profit) during a \cj session of $\Delta t$ minutes, and $h$ be the hash rate of the device in hashes/second. At the time of writing this paper, \ch pays $2,894 \times 10^{-8}$ (XMR; currency unit) for 1 million hashes, where 1 XMR equals \$200 USD. Therefore, the profit $P$ in XMR in $\Delta{t}=  t_f - t_s$ ($t_f$ and $t_s$ refer to the finish and start time of a session, respectively) can be computed as:    
\begin{eqnarray} 
  \label{equ:profit}  P (\text{XMR}) = ({2,894 \times 10^{-8} \times h \times \Delta{t}}) / {10^{6}}
\end{eqnarray} 
The average hash rate of our test device was 21 hashes/second, and for the time  $\Delta{t}=85$ minutes from~\autoref{fig:overviewbattery}, the profit $P$ earned during the session was $3.19 \times 10^{-6}$ XMR or \$ $6.38 \times 10^{-4}$ USD (\$ $1.06 \times 10^{-5}$ USD/second).  This is the upper bound of profit that the device can make in one battery charge. 

To calculate $L$, corresponding to battery drainage due to \cj ($b_n-b_c$), we first measure the time it takes to recharge 1\% of the battery and denote it by $t_r$. Therefore, the time required to recover $b_n-b_c$ can be calculated as $t_r \times (b_n-b_c)$. Let $W$ be the power consumed by the laptop to run for one hour and $C$ be the cost of electricity in USD/KWH. Therefore, the loss $L$ in USD for the use of battery during \cj can be computed using:  
\begin{eqnarray} 
  \label{equ:loss}  L (USD)  = C \times W \times t_r \times (b_n-b_c) 
\end{eqnarray}
For our test device, we had the following parameters: $W=65$ watt-hour, $C = 6.418 \times 10^{-5}$ USD/(watt-hour), $b_n$ = 82\% (in~\autoref{fig:overviewbattery}), $b_c$ = 10\% and $t_r$ = 0.015 hour. Thus, the estimated loss during \cj session $L$ was $\approx$ \$$4.5 \times 10^{-3}$ USD, which is 7 times the value of $P$, highlighting a big gap \cj{}'s operation model. 

Using the same analysis, we examine if \cj can be used as a source of income by users. With the same device as above, the number of hashes required to make 1 XMR (\$$200$ USD) is  $3.45 \times 10^{10}$ hashes. Given that the same device generates 21 hashes/second, the time required to make 1 XMR is approximately 52 years, while the energy consumed is many orders of magnitude more costly (note that the calculations here are quite theoretical; to mine 1 XMR, it would take $\approx$321,543 battery charging cycles, each of which would cost  0.41 cent (total of $\approx 1318$). In~\autoref{tab:eco}, we report all the results obtained from the experiment for each device used in for our experiments in the dynamic analysis, along with the amount of time required for each device to mine 1 XMR. 

\subsection{Cryptojacking and Online Advertisement} \label{sec:mw}

In-browser \cj is being argued as an alternative to online advertisement. To understand the soundness of this argument, we performed an experiment to analyze and compare the monetary value of in-browser \cj as a replacement to online advertisements. 

We select Alexa's top 10 websites~\cite{Alexa-18}. For each website, we obtained the average number of visitors and the time they spent on those websites during March 2018. We use this information and our model from section~\ref{sec:ana} to measure the potential profit those websites could have made using \cj. We assume that visitors on these websites have the average hash rate of 20 hashes/second. We report the results in~\autoref{tab:topsites}, highlighting that those websites would make between \$3.65 million USD (for \textit{youtube.com}) and \$0.10 million USD (\textit{qq.com}) per month (on average). 

Statista~\cite{Statista-17} publishes annual online advertisement revenue reports. We collect the revenues generated by each of those top-10 websites for the year 2017 (most recent report). We use those figures to examine the potential of \cj as an advertisement alternative at scale. For that, we first obtain a monthly revenue figure for each website by dividing the annual revenue by 12. We compare those numbers to the \cj alternative highlighted above. The results are shown in~\autoref{tab:topsites}, where it can be seen that the revenue earned by operating \cj is negligible compared to the revenue earned through online advertisements. For example, if Google is to switch to \cj, it will make \$2.41 million USD per month, at most. In contrast, Google earns $\approx$\$7.94 Billion USD monthly from online advertisement.

To estimate the revenue by \cj websites, we conducted the same experiment on the top-10 websites in our dataset and computed the estimated profit earned by them, shown in~\autoref{tab:cjprofit}. We notice that the maximum profit, earned by \textit{firefoxchina} is $\approx$\$2,747 USD. Although, the ad revenue for these websites is not available online, we still suspect that \$2,747 USD per month is far too low for a website that has 87.24 million monthly views, each with an average duration of 4 minutes and 32 seconds, as compared to the potential revenues for online advertisement. Those findings are in-line with recent reports indicating that an adversary who compromised 5,000 websites and injected his own \cj scripts was only able to make $\$$24 USD~\cite{Hern_18}.

We conclude that in-browser \cj is not a feasible alternative for online advertisement since it generates negligible revenue compared to the existing model. Also, as with most PoW-based systems, the economical analysis of \cj as a model highlights a huge $P$ and $L$ negative gap, making it impractical as a revenue source. 

\begin{table}
\centering
\caption{Monthly Profit earned by top websites by applying cryptojacking. {GR denotes global rank, CR denotes the country rank, visits are in Billions, average time duration of visits is in mm-ss, and P-CJ is profit earned by \cj and P-Ads is revenue earned through ads. ``{---}'' denotes the revenue of the companies that we could not find online.}}
\vs{1}
\label{tab:topsites}
\scalebox{0.9}{
\begin{tabular}{|l|l|l|l|l|l|l|}
\hline
\textbf{Website}  & \textbf{GR} & \textbf{CR} & \textbf{Visits} & \textbf{Time}  & \textbf{P-CJ} &  \textbf{P-Ads} \\ \hline
google.com         & 1                    & 1                     & 47.09                      & 07:23            & 2.41 M  & 7.94 B              \\ \hline
youtube.com           & 2                    & 2                     & 26.22                      & 20:05              & 3.65 M   & 291 M            \\ \hline
baidu.com                  & 3                    & 1                     & 19.08                      & 08:56              & 1.18 M      & 234 M           \\ \hline
wikipedia.org        & 4                    & 6                     & 6.55                       & 03:51              &  0.17 M    & 160 M          \\ \hline
reddit.com            & 5                    & 4                     & 1.69                       & 10:38              & 0.12 M      &  {---}          \\ \hline
facebook.com         & 6                    & 3                     & 29.87                      & 13:28              & 2.80 M      & 3.3 B          \\ \hline
yahoo.com            & 7                    & 7                     & 5.21                       & 06:19              & 0.22 M    & 250 M            \\ \hline
google.co.in               & 8                    & 1                     & 5.33                       & 07:46              & 0.29 M     & 1.1 B           \\ \hline
qq.com                     & 9                    & 2                     & 3.66                       & 04:02              & 0.10 M     &  {---}           \\ \hline
taobao.com                 & 10                   & 3                     & 1.73                       & 06:25              & 0.08 M     &  {---}          \\ \hline
\end{tabular}}
\vs{4}
\end{table}

\begin{table}
\centering
\caption{Estimated monthly earnings of top websites in our dataset. Visits are in millions, average time duration of each visit is in mm-ss and Profit P-CJ is in USD.}
\label{tab:cjprofit}
\scalebox{0.90}{
\begin{tabular}{|l|l|l|l|l|r|}
\hline
\textbf{Website} & \textbf{GR} & \textbf{CR} & \textbf{Visits} & \textbf{Time}  & \textbf{P-CJ} \\ \hline
firefoxchina.cn                & 1,088       & 132        & 87.24             & 04:32      & 2,746.9            \\ \hline
baytpbportal.fi                & 1,613       & 591        & 12.16             & 05:36      & 472.9              \\ \hline
mejortorrent.com               & 1,800       & 37         & 22.83             & 04:50      & 766.4             \\ \hline
moonbit.co.in                  & 2,761       & 1,289      & 15.68             & 28:37      & 3,116.5              \\ \hline
shareae.com                    & 3,331       & 1,071       & 5.86              & 04:49      & 196.0               \\ \hline
maalaimalar.com                & 4,090       & 112        & 3.38              & 03:26      & 80.6            \\ \hline
icouchtuner.to                 & 6,084       & 518        & 7.96              & 02:98      & 200.8            \\ \hline
paperpk.com                    & 6,794       & 2,050      & 3.01              & 03:23      & 70.7               \\ \hline
scamadviser.com                & 6,847       & 668        & 4.20              & 02:08      & 62.2              \\ \hline
seriesdanko.to                 & 7,253       & 1,452      & 5.44              & 04:59      & 188.2             \\ \hline
\end{tabular}}
\vs{2}
\end{table}

\vs{3}
\section{Countermeasures} \label{sec:counter}
In-browser \cj is relatively a new phenomenon therefore, not  much attention has been paid to its use, effects and countermeasures. In this section, we will survey the existing countermeasures available at the browser level to prevent \cj. For the existing countermeasures, we will evaluate their usefulness by performing experiments on our test websites. Furthermore, we will point out new directions for effectively countering \cj based on our results and analysis. 

\subsection{Existing Countermeasures} \label{sec:excm}
At the browser level, existing countermeasures include web extensions such as No Coin, Anti Miner, and No Mining~\cite{Keramidas_18,Tunghobrens-18,Nomining-18}. Each of these web extensions maintains a list of uniform resource locators (URLs) to block while surfing websites. If a user visits a website that is blacklisted by the extension, the user is notified about \cj. However, we show that blacklisting is not an effective technique to counter \cj since an adaptive attacker can always circumvent detection by creating new links that are not found in the public list of blacklisted URLs (proxying). 

To further explore that, we set up these extensions on chrome and evaluated them on our \cj test website. All the extensions detected \cj by reading the WebSocket requests generated by website to \ch. However, in the next phase, we removed the binding key of our script shown in~\autoref{lst:coinhive}. In the absence of the key the website establishes the WebSocket connection but does not perform \cj as it cannot verify itself with the server without the key. However, when we tested that on the extensions, all of them wrongly signaled the presence of active \cj. Since extension-based blacklisting does not read the data frames exchanged between WebSockets, therefore, even the presence of an outdated key or a broken link is falsely labeled as \cj which highlights a major limitation in the detection approach of the existing countermeasures.

\subsubsection{Evading Detection} \label{sec:evadex}
\begin{figure}[]
\begin{center}
\includegraphics[ width=0.45\textwidth]{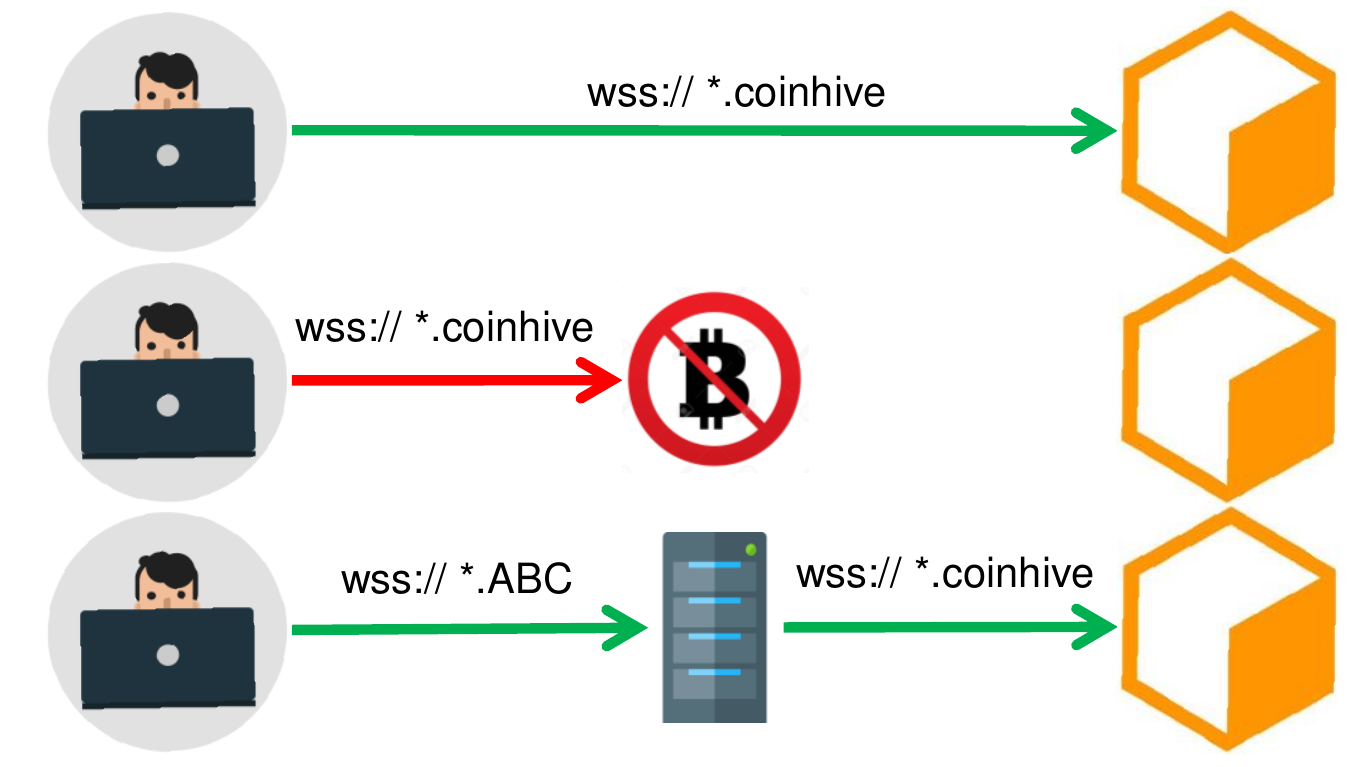}
\caption{Circumventing \cj detection by relaying WebSocket requests through a third party proxy server. }
\vs{10}
\label{fig:circum}
\end{center}
\end{figure}

An attacker, knowing the blacklist, can always evade detection by setting his own third party server to relay data to and from \cj server. The \cj website can establish an innocuous WebSocket connection to a third party server and send data frames and keys to the server. Since anti-\cj extensions will not have the address of third party server blacklisted, they will not be able to prevent the connection and \cj. In~\autoref{fig:circum}, we show how the current countermeasures for \cj can be circumvented by an adaptive attacker. To practically demonstrate that, we set up a test website using \ch script and installed a local relay server. We installed four chrome extensions that block the in-browser \cj, namely No Coin, Anti Miner, No Mining, and Mining Blocker. In the first phase of the experiment, we installed the \ch script and ran the website. Each extension detected the WebSocket request and blocked it. To mimic an adaptive attacker, we configured our relay server to act as a proxy and receive socket requests from the browser and relay them to \ch server. In the \ch script, we modified the code and replaced the \ch socket address with our server address. Next, when we visited the website, it started \cj on the client machine and no extension was able to detect it, concluding it is possible to circumvent detection through a relay server.

\subsubsection{Countering Adaptive Attacker} \label{sec:evadey}
To counter an adaptive attacker and overcome the limitation of existing countermeasures, a better approach is message-based \cj detection in web extensions. Instead of blocking specific URLs, the extensions can monitor the messages exchanged between the user and the server during \cj session. If the messages follow the sequence of web frames that we have illustrated in~\autoref{lst:auth}, the extension can flag them as \cj. This will prevent \cj even if WebSocket requests are relayed through a third party.

To experimentally demonstrate that, we developed a chrome web extension that detects the strings of web frames shown in~\autoref{lst:auth}, and notifies the user when the website starts \cj. To test our extension against the existing countermeasures, we deployed a proxy server that relayed the data between our test website to the {\em dropzone} server as shown in \autoref{fig:circum}. We installed four chrome extensions that detect \cj: No Coin, Anti Miner, No Mining, and Mining Blocker. Since all of these extensions take a blacklisting approach for detection, they failed to detect \cj in the presence of the relay server. However, when we installed our newly developed web extension, it immediately flagged \cj upon reading the actual data exchanged between the browser and the relay server. Therefore, in our view, the blacklisting approach is insufficient to counter \cj. In contrast, better countermeasures can be developed by deeply inspecting the traffic exchanged between the WebSockets.    
\vs{3}

\subsection{Long-term Countermeasures} \label{sec:fd}
Based on the results of our static and dynamic analysis, a fine-grained detection tool can be built at the browser level to address \cj. As we have observed in (\textsection\ref{sec:dynamic}), there are features inherent to the code of \cj scripts that distinguish them form malicious and benign scripts. Moreover, the performance of client machine during \cj is unique in comparison to the performance of the device under normal operation. Based on these features, an accurate detection system can be developed that can detect \cj websites during web browsing. These classifiers can be further used by search engines and web crawlers to identify \cj websites and effectively notify the users about them, or plug them in ``safe-browsing'' lists. 

\subsection{Discussion} \label{sec:discussion} 

By showing a huge negative profit/loss gap, we settle the argument that \cj is not a viable alternative for online advertisement at the moment, and with the current cryptocurrency price. Moreover, the associated negative reputation may also be a factor to discourage users from visiting a website that is known to perform \cj on its visitors. To that end, we do not see browser-based \cj transforming into a popular and ethical way of generating revenues for online web service. This conclusion is also supported by the low prevalence of \cj sites among the top websites in the world, as shown in \autoref{tab:cjprofit}. 

Although the scope of the ethical use of \cj is limited, it is likely that the unethical use may increase as the \cc market grows and the websites remain vulnerable to \js injection attacks. Cryptojacking might not be a suitable revenue source for web service providers, it however, may still provide lucrative incentives for adversaries who can make ``easy money'' by compromising vulnerable websites and targeting their visitors. Malicious website owners may combine both \cj and online advertisements to increase their overall revenue from websites.  

Results from our dynamic analysis (\textsection\ref{sec:dynamic}) show that \cj is highly resource intensive, as it causes excessive battery drainage of the target device. As such, \cj attacks can be launched solely to abuse devices of visitors on a specific website, thereby influencing the reputation of the website and its ability to attract users and traffic. Therefore, \cj provides multiple attack avenues for miscreants and we cannot ignore the potential threat of these attacks or their likelihood of becoming more prevalent in the future. 

As demonstrated in \textsection\ref{sec:evadex}, the existing countermeasures for \cj, based on the blacklisting approach, can be easily circumvented by using relay servers to proxy \cj payload. With the increasing threat potential, and the limitations of defense mechanisms, there is a need for strong countermeasures in the web community to prevent websites from becoming an attack vector for \cj. Web hosting platforms and ISPs can use the methods outlined in our static analysis (\textsection\ref{sec:static}) to keep a check on the spread of \cj code across websites and notify websites' owners and visitors. 

Moreover, as a direct result of our dynamic analysis, we argue that web browsers must shield their users from \cj by analyzing the WebSocket payload (\textsection\ref{sec:dynamic}) and reporting fraudulent behavior to the users. We provide a direction towards such improved countermeasures by developing a chrome extension that reads \cj payload during WebSocket communications, and notifies the users (\textsection\ref{sec:evadey}). With such collaborative efforts and effective defense mechanisms, \cj can be stalled in its early stages from becoming a major threat in future.

\section{Related Work} \label{sec:rw}
In-browser \cj has gained a lot of attention recently, although not treated with any systematic study that covers all major dimensions. In the following, however, we review the related work. 

\subsubsection*{Cryptojacking}
Concurrent to this work, R{\"u}th \etal \cite{RuthWH18} (to be published in ACM IMC 2018; Fall 2018) carried out a measurement study to observe the prevalence of \cj among websites. Towards that, they obtained blacklisted URLs from the No Coin (\textsection\ref{sec:excm}) web extension, and mapped them on a large corpus of websites obtained from the Alexa Top 1M list. In total, they found 1491 suspect websites involved in \cj. However, as shown in \textsection\ref{sec:evadex}, blacklisting approach to detect and prevent \cj has major limitations, and may yield insufficient results to accurately measure prelevance. This perhaps explains a smaller size of their dataset (1491 sites) compared to the dataset used in our analysis (5703 sites). Concurrently, Eskandari \etal \cite{EskandariLMC18} also looked into the prevalence of \cj among websites and the use of \ch as the most popular platform for \cj. Although these two studies, carried out in parallel to ours, highlight the issue of \cj  through measurements, and highlight the emerging use of \cj as an alternative to online ads, they, however, stop short of conducting any code analysis towards detection, nor analyzing the economical arguments for \cj as an alternative online ads, two directions which we pursue in detail in this paper.

Tahir~\etal~\cite{TahirHDAGZCB17} studied the abuse of virtual machines in cloud services for mining digital currencies. They used micro-architectural execution patterns and CPU signatures to determine if a virtual machine in cloud was being illegally used for mining purposes, and proposed \textit{MineGuard}, a tool to detect mining. Bartino and Nayeem~\cite{BertinoN-17} highlighted worms in IoT devices which hijacked them for mining purposes, pointing to the infamous {\em Linux.Darlloz} worm that hijacked devices running Linux on Intelx86 chip architecture for mining~\cite{Bansal-14}. Krishnan~\etal~\cite{KrishnanSV-17} studied a series of computer malware, such as {\em TrojanRansom.Win32.Linkup} and {\em HKTL\_BITCOINMINE}, that turned host machines into mining pools. Sari and Kilik~\cite{SariS-17}, used Open Source Intelligence (OSINT) to study vulnerabilities in mining pools with Mirai botnet as case study.

\subsubsection*{Malicious JavaScript} Malicious \js code and their impact on web browsers and client machines has been studied. Cova~\etal~\cite{CovaKV10} used machine learning techniques to identify anomalous \js code in web applications. Their system also detected obfuscated code, and generated detection signatures for signature-based systems. Classification techniques have been commonly used to detect obfuscated code that appears benign in nature but performs malicious activities~\cite{LikarishJJ09,hou2010malicious,kejriwal2014method}. Jovanovic~\etal~\cite{JovanovicKK06} used static analysis involving context-sensitive data-flow analysis to study vulnerable points Web application programs. Vogat~\etal~\cite{VogtNJKKV07} used dynamic data tainting and static analysis to counter Cross-site scripting (XSS) attacks involving code injection during application launch. Tzermias~\etal~\cite{TzermiasSPM11} combined static and dynamic analysis techniques to detect malicious \js code in vulnerable PDF files. Curtsinger~\etal~\cite{CurtsingerLZS11} presented a \js malware detection tool called ``Zozzle'' that used Bayesian classification and abstract syntax tree
to identify code elements linked to malware.

\subsubsection*{Battery Drain Attacks} Battery is a useful resource in laptops and smart devices, and recently people using smart phones have outnumbered the people using canonical PCs. As a result, the targeted energy-based attacks on smart phone batteries have increased. Fiore~\etal~\cite{FiorePCLS14} studied the energy-based attacks on smart phones and their effect on the battery drainage. Martin~\etal~\cite{MartinHHK04} explored three major attacks namely service request
power attacks, benign power attacks, and malignant power attacks that can be used to drain the battery of pervasive computing devices. A number of other attacks on battery exhaustion have been discovered in mobile phones and laptops that exacerbate the usage of battery sensitive applications and cause swift battery drain~\cite{MoyersDMT10,BuennemeyerGMT07}. 

\vspace{-2mm}
\section{Conclusion} \label{sec:conclusion}
In this paper, we take a systematic look at in-browser cryptojacking through the lenses of characterization, static analysis, dynamic analysis, and economics analysis. In order to that, we collect a dataset of \cj websites and perform static analysis that unveils unique code complexity characteristics and can be used to detect \cj~\js code from malicious and benign code samples with an accuracy of more than 96\%. We explore, through dynamic analysis, how in-browser \cj uses various resources, such as CPU, battery, and network, and use that knowledge to reconstruct the operation of \cj scripts. We also study the economical feasibility of \cj as an alternative to advertising, highlighting its infeasibility. By surveying prior countermeasures and examining their limitations, we highlight long-term solutions, capitalizing on the insights from our static and dynamic analysis, as well as clustering findings. 


\balance
\bibliographystyle{IEEEtran}
\bibliography{references,conf}

\end{document}